\documentclass{article}

\usepackage{arxiv}

\usepackage{times}
\usepackage{latexsym}
\usepackage[T1]{fontenc}
\usepackage[utf8]{inputenc}
\usepackage{microtype}
\usepackage{inconsolata}

\usepackage{amsmath,amssymb,amsthm}
\usepackage{booktabs}
\usepackage{graphicx}
\graphicspath{{figs/}}
\usepackage{xcolor}
\usepackage{algorithm}
\usepackage{algpseudocode}
\usepackage{multirow}
\usepackage{natbib}
\usepackage{url}
\usepackage[hidelinks]{hyperref}
\newtheorem{proposition}{Proposition}
\newtheorem{definition}{Definition}
\newtheorem{remark}{Remark}

\newcommand{\PROCEED}{\textsc{Proceed}}
\newcommand{\VERIFY}{\textsc{Verify}}
\newcommand{\ABSTAIN}{\textsc{Abstain}}
\newcommand{\Vstop}{V_{\mathrm{stop}}}
\newcommand{\Ecal}{\mathbb{E}}

\title{When Is a Memory Safe Enough to Act On?\\
Action-Conditioned Verification for Long-Lived Agents}

\renewcommand{\shorttitle}{When Is a Memory Safe Enough to Act On?}

\author{%
  Yiqi Wang\textsuperscript{1,*}\quad
  Yifei Wang\textsuperscript{2}\quad
  Jiaqi Zhang\textsuperscript{3}\quad
  Zhangkai Wu\textsuperscript{4}\\
  \bfseries Yiqun Duan\textsuperscript{5}\quad
  Taotao Cai\textsuperscript{1}\quad
  Mingkai Zheng\textsuperscript{2}\\[0.6em]
  \normalfont\small
  \textsuperscript{1}University of Southern Queensland, Australia\\
  \normalfont\small\textsuperscript{2}Southern University of Science and Technology, Shenzhen, China\\
  \normalfont\small\textsuperscript{3}Jiangsu University, China\\
  \normalfont\small\textsuperscript{4}The University of Sydney, Australia\\
  \normalfont\small\textsuperscript{5}Uploading Inc., Palo Alto, CA, USA\\[0.4em]
  \normalfont\small\textsuperscript{*}Corresponding author: \texttt{yiqi.wang.jennie@gmail.com}
}

\begin{document}

\maketitle

\begin{abstract}
Long-lived agents act on memory carried across sessions, and many of their actions are
irreversible, so a stale memory can turn directly into harm. Existing work improves the
memory itself --- retrieving, refreshing, or reconciling entries --- yet whether an entry must
be re-checked before acting is not a property of the entry: the same remembered address can
be used unchecked for a routine shipment but should be re-confirmed before a valuable one. We
therefore study \emph{action-conditioned memory verification}: given a candidate action,
decide which of its memory premises to re-verify and when to proceed or abstain, balancing
execution risk, abstention, and verification cost. We formulate this as a sequential stopping
problem, in which a verification is worth its cost only if it can change what the agent does
next; under suitable verifier reliability and cost conditions, increasing action consequence
can move the same memory through proceed, verify, and direct-abstention regimes. Building on
this formulation, \textsc{AMV-Guard} is a practical verification gate that extracts the
premises a candidate action rests on, calibrates beliefs about whether they are wrong, and
selects checks by bounded value of information (VOI). On ToolSandbox, relative to a static
policy given the same inputs, it lowers mean verification cost by 33.9\% while keeping 100\%
safe-proceed recall and 0\% Gate harmful, and raises Proceed coverage from 38.83\% to 44.17\%; in a gate-core transfer to
AppWorld, which carries over the stopping/VOI core but not the full structured pipeline, it
lowers verification cost by 21.3\% at a small cost in safe-proceed recall and task success.
These results suggest that verification effort should be set by what an agent is about to do,
not only by how reliable a memory appears.
\end{abstract}

\section{Introduction}
\label{sec:intro}

Agents that persist across sessions accumulate memory. Much recent work makes that memory
better \citep{packer2023memgpt,xu2025amem} --- retrieving the right
entries, detecting stale ones, tracking provenance, reconciling contradictions
\citep{xu2024knowledge}. All of it answers questions \emph{about the memory}, which
are not sufficient at the moment an agent is about to act --- where recent findings locate
failures. Even when updated evidence has been retrieved, the best model
on a benchmark of invalidated memories reaches 55.2\% accuracy \citep{chao2026stale}, and
agents have been observed to execute an irreversible action and only then recognise that
they should have declined \citep{liu2026agentabstain}.

\begin{figure*}[t]
\centering
\includegraphics[width=\textwidth]{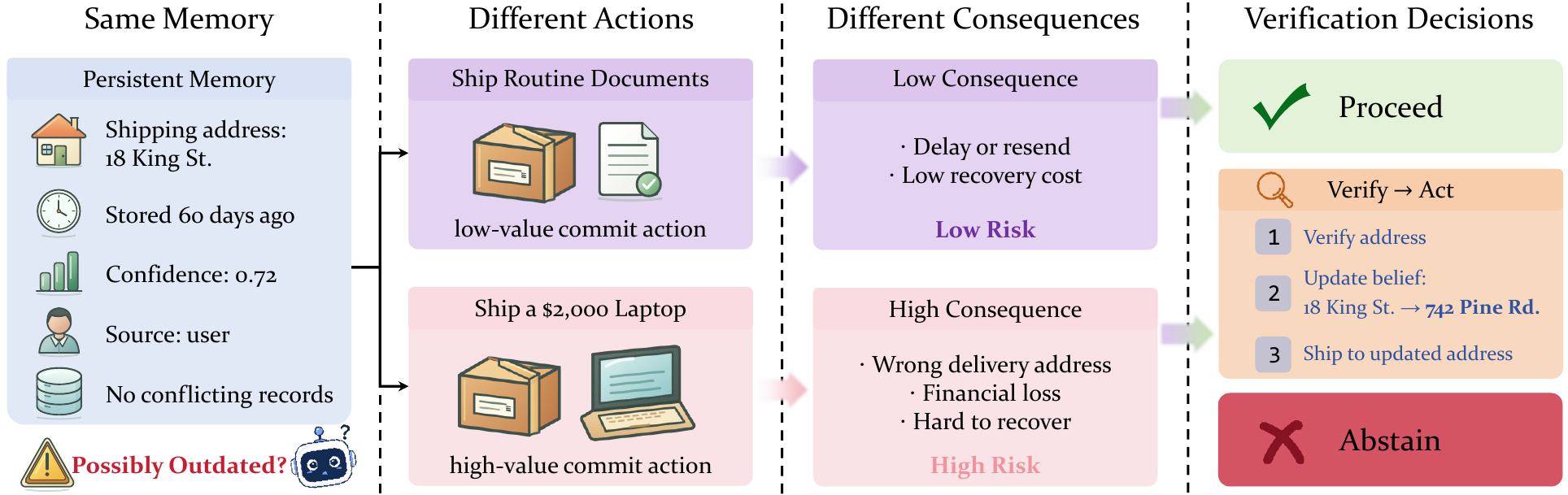}
\caption{Whether a memory should be verified is not a property of the memory. The stored
address, its age, source, and confidence are identical on both rows; only the action and
the cost of being wrong differ, and with them the correct decision.}
\label{fig:motivation}
\end{figure*}

\paragraph{The same memory, different verification needs.}
Consider one entry: \emph{the shipping address is $X$}, stored sixty days ago, from the
user, contradicted by nothing, with a recorded confidence of $0.72$. The agent is about to
ship either routine printed documents or a high-value item to $X$. Every memory-side signal
is identical; the right behaviour is not. In the first case
re-checking costs more than being wrong, so the agent should go ahead. In the second it
should re-confirm first, and if the address cannot be re-confirmed it should decline rather
than guess. One memory, three correct behaviours, selected entirely by the action and the
cost of error (Figure~\ref{fig:motivation}). Whether a memory should be verified therefore
cannot be determined from the memory alone.

\paragraph{Why existing paradigms do not settle it.}
\emph{Memory-centric} methods detect or revise entries that have gone out of date
\citep{chao2026stale}; \emph{environment grounding} methods re-query world-state beliefs
that may have drifted \citep{song2026envprobe} or fetch context an instruction left unstated
\citep{jiang2026accord}; \emph{cost-aware information acquisition} weighs a query against its
price \citep{howard1966information,ding2026cta}. Each contributes something we use, but the
quantity that decides what to check ignores either the action or how several stored
premises jointly support it. An action-independent score
cannot separate the two shipments: for that score they are one instance.

\paragraph{The problem.} We call this \textbf{Action-Conditioned Memory Verification
(AMV)}: given a candidate action $a$, memory graph $G_M$, and verification operations $Q$
with costs and reliabilities, what policy makes executing $a$ sufficiently safe at minimum
total cost --- and when should the agent decline instead? That verification demand depends
on the action is the entry point, not the contribution. Weighing error probability against
asymmetric loss, and paying for information, are well understood, and we build on classical
VOI and Bayesian stopping rather than claim them. What AMV adds is the setting they must
operate in: premises need not be independent, one verifier can cover several, verifiers
differ in cost, reliability, and availability, and each result changes which check, if any,
is worth doing next. A verification that cannot change what the agent
ends up doing has no value, however much uncertainty it removes
\citep{howard1966information}: AMV is a stopping problem, whose value comes from changing the
decision rather than from reducing the expected loss of a fixed action
(Section~\ref{sec:voi-fix}).

\paragraph{Contributions.}
\begin{enumerate}\itemsep2pt
\item \textbf{Problem and formalisation.} We distinguish memory validity from
action-specific verification demand --- separating \emph{relevant},
\emph{action-supporting}, and \emph{critical} memories --- and cast verification as a
stopping problem with abstention as a terminal decision. An explanatory proposition shows
that, under suitable verifier reliability and cost conditions, increasing consequence can
move one memory through proceed, verify, and decline-\emph{without}-verifying regimes
(Section~\ref{sec:formulation}).
\item \textbf{Method.} \textsc{AMV-Guard} combines premise extraction, calibrated
memory-error beliefs, and bounded VOI selection (greedy one-step VOI with a two-step
fallback) into an executable verification gate (Section~\ref{sec:method}).
\item \textbf{Empirical findings.} We measure the gate's verification-cost gains, analyse
when adaptive verification helps, and report transfer trade-offs and failure boundaries,
with an information-constrained exact planner as a regret reference, not supervision (Sections~\ref{sec:protocol} and~\ref{sec:experiments}).
\end{enumerate}

\section{Problem Formulation}
\label{sec:formulation}

\subsection{Setup}

An agent carries persistent memory claims $M=\{m_1,\dots,m_n\}$, each with provenance
(write time, source, supporting observation, recorded confidence), held in a memory graph
$G_M$. The world is in a hidden state $z\in\mathcal{Z}$; the agent sees only $G_M$, possibly
out of date, and holds a belief $b$ over $z$ induced by $G_M$ and a model of how claims fail.

The planner proposes a \emph{candidate action} $a$. Only \textsc{Commit} actions,
which change state or have external consequence, reach the gate. A commit rests on
\emph{premises} $P(a)$: propositions about $z$ that must hold for $a$ to be executable,
correctly parameterised, and permitted. A verification $q\in Q$ costs $C(q)$ and returns
$o$ with likelihood $\Pr(o\mid z,q)$. Verifiers are not in bijection with premises: one
query can resolve several, and some premises have none. Terminal decisions are
candidate-local, $\mathcal{D}(a)=\{\PROCEED(a),\ABSTAIN\}$, with loss $L(d,z)$ asymmetric
across both arguments. Replanning is not a terminal action: if evidence invalidates a required
premise, the Gate rejects $a$ and the agent loop may submit a new candidate.

\subsection{Three notions the literature merges}

\begin{definition}[Relevant / action-supporting / critical]
$m$ is \emph{relevant} if semantically related to the task --- this is what retrieval
returns. $m$ is \emph{action-supporting} for $a$ if it supports some premise in $P(a)$, so
the correctness of $a$ is derived through $m$. $m$ is \emph{critical} for $(a,L)$ if an
error in it materially changes the loss of executing $a$: formally, if replacing $m$'s
asserted value by its true value changes $L(\cdot,z)$ for the executed action.
\end{definition}

These nest: critical $\subseteq$ action-supporting $\subseteq$ relevant. Retrieval
returns the outer set; verification budget should be spent on the inner one.
Criticality is a property of $(a,L)$, while every memory-side feature is a property of $m$
alone.

\begin{proposition}[Consequence induces three regimes]
\label{prop:regimes}
Consider the expected-loss special case without the additional $\delta,\epsilon$ execution
screens. Fix a belief assigning probability $p\in(0,1)$ to a premise being false, an abstention cost
$\kappa>0$, and one verifier of cost $c\ge 0$ with symmetric reliability
$r\in(\tfrac12,1]$. Let executing under a false premise cost $\theta>0$, all else held
fixed, and write $\pi=\Pr(\mathrm{confirm})$ and
\begin{gather*}
p^{+}=\frac{p(1-r)}{(1-p)r+p(1-r)},\\
p^{-}=\frac{pr}{(1-p)(1-r)+pr}
\end{gather*}
for the posteriors after a confirming and a refuting observation. Then
$p^{+}<p<p^{-}$, and there are thresholds $\theta_1\le\theta_2$ such that the optimal
initial decision is \PROCEED\ for $\theta<\theta_1$, \VERIFY\ for
$\theta_1\le\theta<\theta_2$, and \ABSTAIN\ for $\theta\ge\theta_2$. The verification
region is bracketed by
\[
\kappa/p^{-}\;\le\;\theta_1\;\le\;\theta_2\;\le\;\kappa/p^{+},
\]
it is empty when $c$ is large, and on $\theta\ge\kappa/p$ verification is chosen exactly
when $c<\pi\,(\kappa-\theta p^{+})$. For $r<1$, $\theta_2$ is finite. For $r=1$,
$\theta_2=\infty$ only when $c<(1-p)\kappa$; otherwise the high-consequence decision is
direct abstention and the verification region is empty.
\end{proposition}

\noindent Below $\theta_1$ verification is not worth its cost; below $\kappa/p^{-}$ its gross
value is exactly zero, by the mechanism of Remark~\ref{rem:degenerate}. The regimes are traversed by varying $\theta$ with the memory held fixed: a score computed
from $m$ and $G_M$ alone is constant in $\theta$ while the optimal decision is not. With an
imperfect verifier there is a consequence level past which even a confirming observation
cannot license execution, so the rational act is to decline \emph{without} verifying. With a
perfect verifier, verification remains optimal at arbitrarily high consequence when
$c<(1-p)\kappa$; otherwise, verification is never strictly preferred.
Hence \ABSTAIN\ must be a terminal decision the policy can reach directly, and verifier
reliability must be modelled. Proof and extended discussion are in Appendix~\ref{app:proof}.

\subsection{The decision problem}
\label{sec:decision-problem}

For a candidate $a$, define its posterior expected execution loss and posterior harm
probability as
\begin{equation}
\begin{aligned}
R(b,a)&=\Ecal_{z\sim b}[L(\PROCEED(a),z)],\\
H_\delta(b,a)&=\Pr_{z\sim b}\!\left(L(\PROCEED(a),z)>\delta\right).
\end{aligned}
\label{eq:risk-harm}
\end{equation}
where $\delta$ is an expected-loss threshold and $\epsilon$ a posterior harm-probability
limit. With abstention cost $\kappa$, the candidate-local stopping value implemented by the
Gate is
\begin{equation}
\Vstop(b,a)=
\begin{cases}
R(b,a), & \text{if } R(b,a)\le\min\{\delta,\kappa\}\\
        & \text{and } H_\delta(b,a)\le\epsilon,\\
\kappa, & \text{otherwise.}
\end{cases}
\label{eq:vstop}
\end{equation}
\PROCEED{} is admitted only when both risk screens hold and it is no more costly than
abstaining; the screens decide admissibility, and VOI decides whether further information is
worth its cost (Appendix~\ref{app:formulation}). With remaining budget $B$ and posterior
$b^{q,o}$ after running $q$ and observing $o$, the information-constrained exact reference
satisfies
\begin{multline}
V^*(b,B,Q;a)=\min\Bigl\{\Vstop(b,a),\\
\min_{q\in Q:\,C(q)\le B}\bigl[\lambda C(q)\\
+\Ecal_o V^*(b^{q,o},B-C(q),Q\setminus\{q\};a)\bigr]\Bigr\}.
\label{eq:bellman}
\end{multline}
Equation~\eqref{eq:bellman} is an evaluation reference, not the deployed policy; the screen
$H_\delta\le\epsilon$ is a per-candidate decision rule, distinct from the empirical Gate
harmful metric.

\subsection{Why expected loss reduction for a fixed action is insufficient}
\label{sec:voi-fix}

It is natural to score a verification by how much it reduces the risk of the candidate
action: writing $R(b,a) = \Ecal_{z\sim b}[L(a,z)]$, this gives $\mathrm{VOI}(q) = R(b,a) -
\Ecal_o[R(b^{q,o},a)] - \lambda C(q)$. It is degenerate.

\begin{remark}[The naive objective is never positive]
\label{rem:degenerate}
For a fixed action $a$, when verification yields information without changing the world
state and posteriors are formed by Bayes' rule,
\begin{align*}
\Ecal_o\bigl[R(b^{q,o},a)\bigr]
&= \sum_z b(z) L(a,z) \sum_o \Pr(o \mid z)\\ &= R(b,a),
\end{align*}
so $\mathrm{VOI}(q) \equiv -\lambda C(q) \le 0$ for every $q$, strictly negative whenever
$\lambda C(q)>0$. A planner using this score
with the stopping rule ``verify while some $\mathrm{VOI}$ is positive'' never verifies.
\end{remark}

Observations do not lower the average loss of a \emph{fixed} action; information has value
only through the ability to change the terminal decision \citep{howard1966information}. Replacing fixed-action risk by the screened stopping value
gives the implemented one-step score:
\begin{multline}
\mathrm{VOI}_1(q;b,B,Q,a)=\Vstop(b,a)\\
-\left[\lambda C(q)+\Ecal_o\Vstop(b^{q,o},a)\right].
\label{eq:voi}
\end{multline}
Because the screens make Equation~\eqref{eq:vstop} piecewise, VOI is computed directly under
the verifier observation model.

One-step VOI misses complementary checks: neither query alone need change the stopping
decision even though the pair does. Let $W_1(b,B,Q;a)$ be the best value of stopping
immediately or taking one further affordable verifier $r$, i.e.\
$\min\{\Vstop(b,a),\min_{r}[\lambda C(r)+\Ecal_{o'}\Vstop(b^{r,o'},a)]\}$. The bounded
two-step score for a first query $q$ is then
\begin{multline}
\mathrm{VOI}_2(q;b,B,Q,a)=\Vstop(b,a)-\bigl[\lambda C(q)\\
+\Ecal_o W_1\!\left(b^{q,o},B-C(q),Q\setminus\{q\};a\right)\bigr].
\label{eq:voitwo}
\end{multline}
This is a depth-two lookahead, not the full Bellman policy.

\section{AMV-Guard}
\label{sec:method}

\begin{figure*}[t]
\centering
\includegraphics[width=\textwidth]{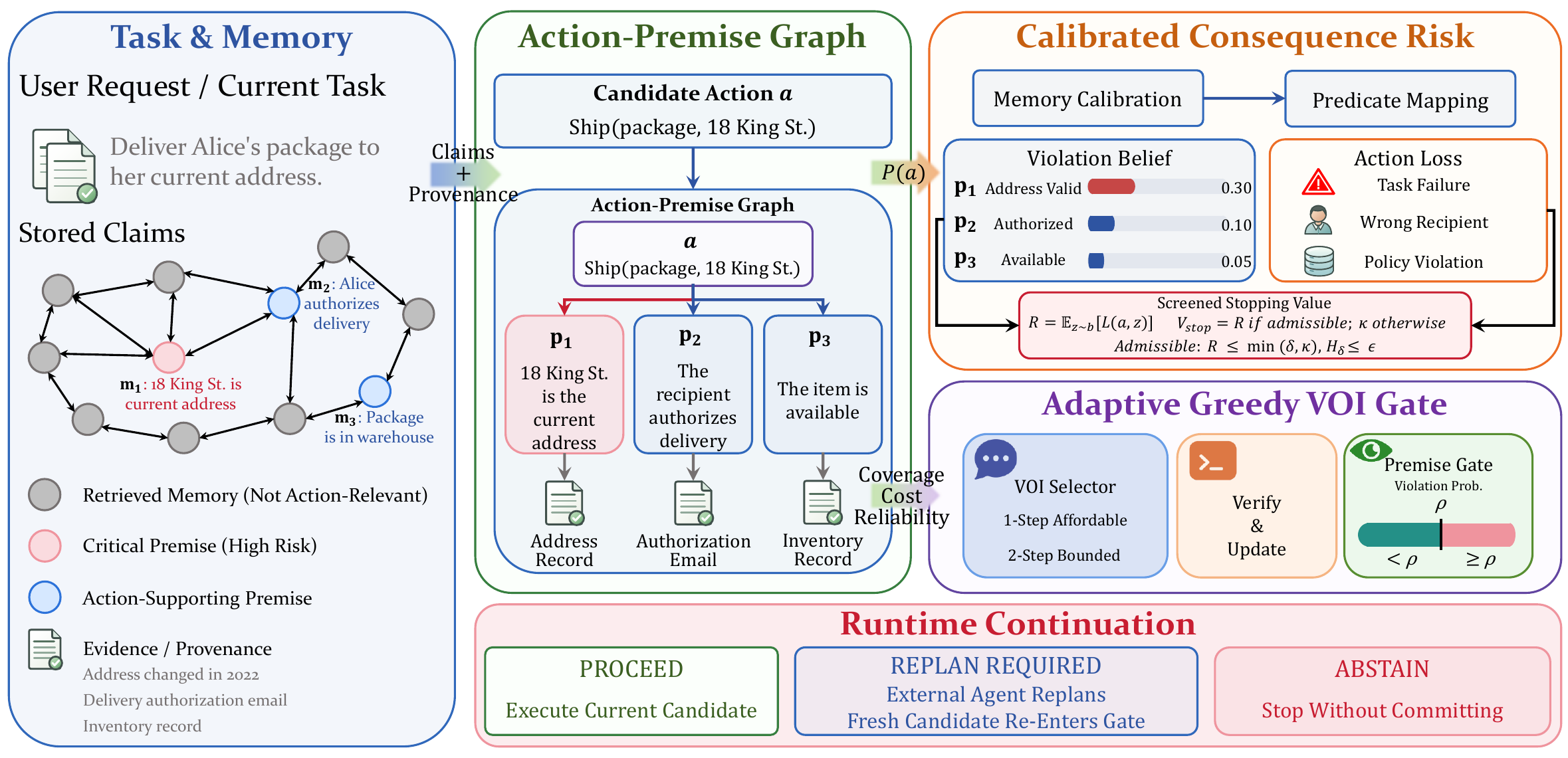}
\caption{\textsc{AMV-Guard}. Retrieved memory is narrowed to the premises the candidate
rests on; calibrated premise beliefs are combined with an action-dependent loss; the policy
applies the screened stopping value and adaptive VOI, re-deciding after each observation. An
invalidated premise returns the candidate to the agent loop; replanning is not a VOI
terminal action.}
\label{fig:pipeline}
\end{figure*}

\textsc{AMV-Guard} sits between planner and executor: for each proposed commit it identifies
the premises, forms calibrated beliefs over them, and verifies, proceeds, or abstains
(Figure~\ref{fig:pipeline}).

\subsection{From retrieved memories to action premises}
\label{sec:premises}

From the candidate action, its tool schema, the plan prefix, and the retrieved memory graph
we extract candidate premises, each tagged with its supporting claims and a dependency type:
direct (the premise enters the tool arguments), transitive, or joint. An LLM extractor and a
public tool-contract rule layer propose premises independently. LLM proposals must ground to
a memory identifier, verifier, or public action target; the rule layer adds required
arguments and environment and policy preconditions. A counterfactual filter learned on the training split removes
repeatedly irrelevant tool--target pairs; hidden dependency labels are never inference
inputs. The current implementation uses dependency tags and simplified coupling assumptions
rather than a general joint inference model over arbitrary premise dependencies: the risk
engine keeps verifier-to-premise coverage sets but represents a multi-premise verifier by one
shared binary observation, and premises tagged joint share one latent draw, which
approximates rather than performs joint dependency inference (Appendix~\ref{app:method}).

\subsection{From memory confidence to calibrated, consequence-aware risk}
\label{sec:calibration}

Model-reported confidence is informative \citep{kadavath2022language,tian2023just} but
overconfident \citep{xiong2024express} and unreliable when a belief is confidently wrong
\citep{song2026envprobe}, so we use it as one feature rather than as a probability. For each
action-supporting claim we estimate $p_i = \Pr(\text{claim incorrect}\mid\text{features})$
from features spanning recency, source, consistency, graph position, and model signals
(Appendix~\ref{app:features}). We fit a logistic base model on training instances and select
among Platt, beta, and bounded-isotonic calibration using held-out-family likelihood and
calibration error; development data selects decision hyperparameters and test data is
evaluation-only \citep{zadrozny2002transforming,guo2017calibration}. Memory-error
probabilities are converted into action-precondition violation probabilities
(Appendix~\ref{app:method}), which induce a belief over $z$, and $L$ turns it into the
stopping value of Equation~\eqref{eq:vstop}. The asymmetry lives entirely in $L$: the
ToolSandbox instantiation maps premises to TaskFailure, WrongRecipient, PolicyViolation, or
OtherFailure, fits the first three weights from the training split, and charges each loss
group once.

\subsection{Adaptive greedy verification with a two-step fallback}
\label{sec:policy}

At every Gate step the policy
computes Equation~\eqref{eq:voi} for every affordable verifier and takes the largest positive
score. If none is positive and two-step lookahead is enabled, it takes a positive maximiser
of Equation~\eqref{eq:voitwo} as the first query of a potentially complementary pair. If
neither stage has positive value, it returns the terminal decision from
Equation~\eqref{eq:vstop}. After each observation the posterior is updated by Bayes' rule and all
scores recomputed: a bounded receding-horizon approximation to
Equation~\eqref{eq:bellman}, not a claim of globally optimal planning.

\paragraph{Stopping and replanning.}
The Gate returns \PROCEED{} only when Equation~\eqref{eq:vstop} admits execution and no
positive-VOI query remains, and \ABSTAIN{} when no query has positive value and execution is
inadmissible --- for example because a critical premise is unverifiable, available verifiers
are too noisy, or the budget is exhausted. A positive abstention cost
$\kappa = L(\ABSTAIN,\cdot)$ penalises unnecessary refusal. If the posterior probability that a required premise is violated reaches the
invalidation threshold $\rho$, the Gate returns \textsc{ReplanRequired} with the changed
premise and evidence instead of executing the stale candidate. We select $\kappa$,
$\lambda$, $\rho$, and the two-step switch on development data under a safety-first rule.
Algorithm~\ref{alg:main} (Appendix~\ref{app:alg}) gives the full loop;
Appendix~\ref{app:method} gives replanning details and computational cost.

\section{Evaluation Protocol and Oracles}
\label{sec:protocol}

\subsection{Replayable instances}

A history session runs from a clean initial state, writing to memory only facts the agent
observed through a tool or the user supplied --- simulator ground truth is never injected.
The state is snapshotted and a hidden perturbation applied, drawn from ten families spanning
no change, single and multiple critical changes, irrelevant and correlated changes,
unverifiable premises, noisy and contradictory verifiers, cost-inverted verifier sets, and
misleading source trust (Appendix~\ref{app:perturbations}). The agent then runs on memory
alone until its first commit, which is frozen as the candidate. Environment state and
verifier observations replay exactly from snapshot and seed; all variants of a base scenario
share a split.

\subsection{Candidate-fixed safety interventions}
\label{sec:interventions}

Starting from a reference state satisfying the proposed preconditions, we execute the frozen
candidate, flip each candidate premise individually, and mark it as a direct dependency when
the execution outcome changes. For premise pairs whose individual flips do not change the
outcome, we test the pair and record a pairwise joint dependency when the joint intervention
does. These labels determine whether the safety of the frozen action depends on a state
premise; we do not re-run the planner and do not use planner-reexecution labels as Gate
supervision.

\subsection{Two oracles, and what each measures}
\label{sec:oracle-policy}

\paragraph{Perfect-information structural target oracle.} Given the hidden
post-perturbation state, this oracle identifies stale or missing action dependencies and
the logical premise targets that would need perfect observation. It is used for instance
auditing, extraction scoring, and coverage analysis, not as a decision reference.

\paragraph{Information-constrained exact planner.} An unperturbed instance and one with a
single critical change present the agent with identical information before the first
verification, so any policy respecting the agent's information set must behave identically
on both at that step. Regret to labels derived from hidden truth mixes decision regret
with irreducible uncertainty. We therefore solve
Equation~\eqref{eq:bellman} with the information available to the deployable Gate ---
predicted dependencies, calibrated beliefs, verifier costs, reliabilities, and observation
model, but not hidden truth. This \emph{ExactPlanner-PredictedDep} is a decision-agreement
and regret reference; it does not supervise the main method. Oracle-dependency and
oracle-belief substitutions are diagnostic upper bounds, not deployable variants.
Determinism gives replay and exactness within the finite modelled problem, not premise
completeness, calibration correctness, or observation-model correctness
(Appendix~\ref{sec:oracle-limits}).

\section{Experiments}
\label{sec:experiments}

\subsection{Setup}
\label{sec:setup}

\paragraph{Environments.} Two executable environments
\citep{lu2025toolsandbox,trivedi2024appworld} with native state-based evaluators, so
success, policy violations, and unintended state changes are measured rather than judged by
a language model. ToolSandbox contains 3,520 frozen instances split into 2,240/640/640
train/development/test instances; the main test is repeated with three observation seeds.
AppWorld, used for a gate-core transfer with recalibration on its own development split and
no retraining, contains 100 state-mutating tasks under clean and stale-memory conditions
(200 episodes).

\paragraph{Held fixed across methods.} Within each benchmark we hold fixed the action agent
and backbone, verifier interfaces, costs and reliabilities, episode budget, replanning
prompt, and perturbation seeds. ToolSandbox candidates are frozen from the same ungated
rollout. AppWorld instead intercepts commits from its native agent loop and maps each to one
compound required-state premise plus one compound verifier, so it transfers the calibrated
stopping/VOI core, not the premise extractor, multi-verifier selection, or two-step
fallback.

\paragraph{Baselines and references.} The main comparisons include always proceed,
always verify, static AMV, and adapted environment-probing and cost-aware baselines
\citep{song2026envprobe,ding2026cta}; their exact inputs and thresholds are listed in
Appendix~\ref{app:baselines}. Where an information-matched variant is implemented, it
receives the same calibrated beliefs, predicted dependencies, costs, and reliabilities as
\textsc{AMV-Guard}. 

\paragraph{Metrics, and why coverage is reported everywhere.} Harmful action rate and
verification cost can both be driven to zero by declining everything. Safe-proceed recall
is the fraction of ground-truth-safe, Gate-evaluable original candidates executed during
the initial Gate episode, possibly after verification and before replanning; the harmful-candidate
analogue, \emph{Unsafe Pass Rate}, is reported as a diagnostic. The primary \emph{Gate harmful}
metric uses the same harmful-execution numerator over all Gate-evaluable instances. On
ToolSandbox, these metrics exclude the 40 test instances with initial-candidate action or
intent errors, leaving 600 per seed and 1,800 over three seeds. Proceed coverage is the
fraction of Gate-evaluable instances in which any Gate-approved action executes, including
after replanning. Instances that finish after feedback without such an execution are labelled
\emph{Resolved without action} and count as neither Proceed nor ABSTAIN. Final ABSTAIN likewise
requires no Gate-approved execution, an explicit terminal ABSTAIN, and no resolution without action.
Task success is end-to-end; raw harm records any executed action whose observed loss exceeds
the threshold, including planning or intent failures. Verification cost accumulates across
all Gate entries in an episode. ToolSandbox uncertainty uses intervals across three observation seeds;
AppWorld uses paired task-level bootstrap intervals with 2,000 resamples.

\subsection{RQ1: Does action-conditioned verification improve the safety--cost trade-off?}
\label{sec:rq1}

\begin{table*}[t]
\caption{ToolSandbox live end-to-end results (3 observation seeds $\times$ 640 test
instances; intervals in Table~\ref{tab:toolsandbox-ci}). Metric definitions in
Section~\ref{sec:setup}. Bold: best per column, excluding Always proceed/verify and the exact planner.}
\label{tab:main}
\centering
\footnotesize
\resizebox{\textwidth}{!}{%
\begin{tabular}{lcccccc}
\toprule
Method & Safe recall $\uparrow$ & Gate harmful $\downarrow$ & Verify cost $\downarrow$ & Task success $\uparrow$ & Raw harm $\downarrow$ & Coverage $\uparrow$ \\
\midrule
Always proceed        & 100.00\% & 66.67\% & 0.000 & 68.23\% & 68.75\% & 100.00\% \\
Always verify all     & 100.00\% & 23.67\% & 1.976 & 74.58\% & 28.02\% & 62.67\% \\
EnvProbe-Simple       & \textbf{100.00\%} & 49.00\% & 2.018 & \textbf{74.48\%} & 52.19\% & 87.00\% \\
CTA                   & 99.83\% & 63.61\% & \textbf{0.123} & 67.71\% & 65.89\% & \textbf{96.89\%} \\
AMV-Static            & \textbf{100.00\%} & \textbf{0.00\%} & 1.843 & 72.34\% & \textbf{6.82\%} & 38.83\% \\
\textsc{AMV-Guard} (ours) & \textbf{100.00\%} & \textbf{0.00\%} & 1.219 & 73.91\% & 7.03\% & 44.17\% \\
\midrule
\emph{ExactPlanner-PredictedDep} & 100.00\% & 0.00\% & 1.219 & 73.91\% & 7.03\% & 44.17\% \\
\bottomrule
\end{tabular}
}
\end{table*}

Table~\ref{tab:main} reports the comparison; Figures~\ref{fig:safety-coverage}
and~\ref{fig:loss-cost} (Appendix~\ref{app:results}) sweep the operating point. Across the safe $\delta\le0.5$ operating points, measured Gate harmful is
0\%, while coverage ranges from 0\% to 46.50\% and mean verifier calls from 0 to 1.153. The
main point has 44.17\% Proceed coverage, 73.91\% task success, 7.03\% raw end-to-end harm,
and 1.219 mean verification cost; under the current modelling assumptions it coincides with
ExactPlanner-PredictedDep on every reported metric, which is agreement within the modelled
problem rather than evidence of general optimality. Relative to AMV-Static, it reduces cost by
33.9\% while both methods retain 100\% safe-proceed recall, 0\% Unsafe Pass, and 0\% Gate
harmful. This saving accompanies 5.33 points more Proceed coverage and 1.56 points more task
success, with raw harm 0.21 points higher. Relative to verify-all, it reduces cost by 38.3\%.

Across Qwen3.5-35B-A3B, Qwen3.5-27B, and DeepSeek-V4.1-Flash, each
with its own candidates, calibration, and thresholds, adaptive VOI reduces mean verification cost for all
three backbones, by 33.9\%, 42.3\%, and 58.9\% (Tables~\ref{tab:cross-model-compact}
and~\ref{tab:cross-model-full}). Both Static and adaptive policies have zero measured Gate
harm for all three backbones. Coverage increases by 5.33 points for Qwen3.5-35B-A3B, is
unchanged for Qwen3.5-27B, and decreases by 5.68 points for DeepSeek-V4.1-Flash.
Table~\ref{tab:cross-model-full} also reports the less
favourable downstream outcomes.

\subsection{RQ2: What do action conditioning, consequence, and premise structure each contribute?}
\label{sec:rq2}

\paragraph{Action conditioning.} A paired intervention ($n=500$ profiles) fixes the memory
graph (contact, cellular, Wi-Fi, and low-battery claims), impacts, verifiers, and Gate
parameters, and changes only the candidate: \texttt{send\_message\_with\_phone\_number}
selects the contact and cellular premises, \texttt{set\_wifi\_status} the other two, with
error probability $0.30$ on one action's premises and $0.01$ on the other's. The high-risk action verifies and the
low-risk action proceeds in 100\% of profiles, and every selected query covers a premise of
its own candidate (Table~\ref{tab:action-conditioning}).

\paragraph{Consequence.} Holding the action and memory evidence fixed and rescaling only
total loss ($n=500$ per level), AMV verifies 0\%, 15.8\%, and 62.8\% of low-, medium-, and
high-loss cases; its high-loss harmful-proceed rate is 0\%, versus 64.2\% without loss
conditioning. Baselines without loss conditioning verify at the same rate at every level, so their
high-loss harmful-proceed rates stay at 28.2--66.4\% (Table~\ref{tab:consequence}).

\paragraph{Premise extraction and calibration.} Extraction is evaluated under
distribution shift in Section~\ref{sec:rq4}. Calibration lowers ECE/Brier/NLL from 0.319/0.319/4.401 for raw
confidence to 0.082/0.209/0.588 with all fitting data, at which the Gate retains 100\%
safe-proceed recall and 0\% Gate harmful (Table~\ref{tab:calibration-data},
Figure~\ref{fig:calibration-reliability}).

\subsection{RQ3: When does adaptive multi-step verification help?}
\label{sec:rq3}

Static and adaptive verification coincide on many simple ToolSandbox candidates, so a
controlled subset adds more premises, heterogeneous verifier costs and
reliabilities, overlapping coverage, and cases requiring more than one observation
(Table~\ref{tab:adaptive}); it uses repeated one-step greedy selection without the two-step
fallback. Relative to AMV-Static, adaptivity lowers expected objective by 0.0314 (95\% CI
$[-0.0322,-0.0306]$), increases safe completion by 4.15 percentage points, and reduces regret
to the exact planner from 0.0746 to 0.0431, at 0.257 additional mean verifier cost. Its
6.8\% multi-step rate shows that the gain is concentrated in these controlled conditions
rather than universal. At the selected ToolSandbox development configuration
($\kappa=1$, $\lambda=0.1$, $\rho=0.95$), enabling the two-step fallback raises
safe-proceed recall from 17.0\% to 100\% and Proceed coverage from 5.31\% to 31.25\%, while
both settings retain 0\% harmful proceeds; mean verification cost rises from 0.059 to 1.213.
This supports a bounded adaptive fallback, not dominance of adaptivity on every benchmark.

\begin{table*}[t]
\caption{Controlled adaptive subset ($n=2{,}000$). Multi-step: share using more than one
verifier; objective: decision loss plus weighted verification cost. Rates are percentages.
Bold as in Table~\ref{tab:main}.}
\label{tab:adaptive}
\centering
\scriptsize
\resizebox{\textwidth}{!}{%
\begin{tabular}{lrrrrrr}
\toprule
Policy & Harmful $\downarrow$ & Cost $\downarrow$ & Multi-step & Safe completion $\uparrow$ & Objective $\downarrow$ & Regret $\downarrow$ \\
\midrule
Always Proceed & 14.20 & 0.000 & 0.00 & 85.80 & 0.6824 & -- \\
Always Verify & 0.55 & 14.250 & 100.00 & 84.05 & 0.8903 & -- \\
AMV-Static & \textbf{1.10} & \textbf{3.750} & 0.00 & 77.35 & 0.4550 & 0.0746 \\
AMV-Adaptive (one-step) & 1.20 & 4.007 & 6.80 & \textbf{81.50} & \textbf{0.4236} & \textbf{0.0431} \\
Exact Planner & 1.70 & 2.547 & 38.35 & 81.40 & 0.3805 & 0.0000 \\
\bottomrule
\end{tabular}}
\end{table*}

\subsection{RQ4: Does it hold away from training conditions?}
\label{sec:rq4}

AppWorld changes the environment and agent loop; its development split is used only for
recalibration, and we do not treat it as a full-pipeline migration. On this transfer (Table~\ref{tab:appworld}),
AMV lowers verification cost by 21.28\% relative to AMV-Static (2.090 vs.\ 2.655) and raw
harm by 0.50 percentage points, with 7.89\% vs.\ 8.42\% Gate harmful, but does not improve
task success (28.00\% vs.\ 30.50\%) or safe-proceed recall (70.53\% vs.\ 72.63\%). Proceed
coverage is 57.37\%, and bootstrap intervals are broad (Table~\ref{tab:appworld-ci}). This
supports transfer of the decision core, not yet of the full structured pipeline. On the controlled
premise-extraction strata (160 candidates each; Table~\ref{tab:premise-ood}), the deployed
Union+CF extractor (LLM+Graph with counterfactual filtering) has critical recall 95.91\%, 92.45\%, and 94.97\% and Gate harmful
8.23\%, 15.19\%, and 10.13\% on unseen tool families, unseen domain schemas, and unstructured
language, with dependency accuracy falling to 61.92\% on unstructured language; its safe-proceed
recall remains 100\% in all four strata.

\begin{table*}[t]
\caption{AppWorld gate-core transfer (100 tasks, 200 clean/stale episodes per method; 190
Gate-evaluable). Rates are percentages. Bold as in Table~\ref{tab:main}.}
\label{tab:appworld}
\centering
\scriptsize
\resizebox{\textwidth}{!}{%
\begin{tabular}{lrrrrrr}
\toprule
Method & Safe recall $\uparrow$ & Gate harmful $\downarrow$ & Cost $\downarrow$ & Task success $\uparrow$ & Raw harm $\downarrow$ & Coverage $\uparrow$ \\
\midrule
Always Proceed & 100.00 & 18.95 & 0.000 & 35.50 & 19.50 & 100.00 \\
Always Verify & 54.74 & 6.32 & 5.600 & 11.00 & 8.50 & 37.37 \\
EnvProbe & \textbf{91.58} & 17.37 & 2.980 & 23.00 & 18.50 & \textbf{90.53} \\
CTA & 67.37 & 13.16 & 3.385 & 18.50 & 16.00 & 61.05 \\
AMV-Static & 72.63 & 8.42 & 2.655 & \textbf{30.50} & 10.50 & 59.47 \\
\textsc{AMV-Guard} & 70.53 & \textbf{7.89} & \textbf{2.090} & 28.00 & \textbf{10.00} & 57.37 \\
\bottomrule
\end{tabular}}
\end{table*}

\subsection{RQ5: Where does it fail?}
\label{sec:rq5}

Separating failures by ownership: on ToolSandbox, \textsc{AMV-Guard} has 0\% Unsafe Pass and
0\% Gate harmful, while raw end-to-end harm is 7.03\% once planning/intent failures are
included; 6.25\% of episodes have an initial-candidate action or intent error, 31.83\% of
Gate-evaluable instances are resolved after feedback without a Gate-approved execution, and
25.00\% invoke replanning
(Table~\ref{tab:toolsandbox-secondary}). On AppWorld, 7.89\% Gate harmful (15.79\% Unsafe Pass) remains,
and gate-local and end-to-end metrics need not move together.

\section{Related Work}
\label{sec:related}

\paragraph{Long-term agent memory.} Much work improves what an agent stores and retrieves:
relevance and consolidation \citep{packer2023memgpt,zhong2024memorybank,gutierrez2024hipporag,
xu2025amem}, experience reuse \citep{park2023generative,shinn2023reflexion}, outdated
knowledge \citep{vu2024freshllms}, stale-memory detection and write-time revision
\citep{chao2026stale}, knowledge conflicts \citep{xu2024knowledge,xie2024adaptive}, and, for
knowledge held in model weights, direct editing \citep{meng2022locating}. Proactive
variants decide \emph{when to surface} a memory \citep{wu2026remember}. These improve the
store, but the quantities they compute are action-independent --- the gap
Proposition~\ref{prop:regimes} formalises and Section~\ref{sec:rq2} measures.

\paragraph{Active grounding and environment probing.} ACCORD probes the environment
for context an instruction left unstated \citep{jiang2026accord}. EnvProbe, our primary
comparison, queries one belief field per step of a structured world model, scored by task
criticality, staleness, verbalised uncertainty, and dependency role
\citep{song2026envprobe}, and likewise treats calibration as action selection. We differ: we verify \emph{persistent} memory claims
with provenance rather than a current belief table; our verifiers differ in cost and
reliability and some premises have none; and we select sequences, model multi-premise verifiers, and let the agent decline.

\paragraph{Cost-aware acquisition and selective prediction.} Weighing a query against its
price and knowing when to stop is classical \citep{howard1966information,wald1945sequential,
golovin2011adaptive}; recent agent work applies it before committing \citep{ding2026cta}; and
abstention supplies the third terminal decision, from the classical reject option
\citep{chow1970optimum} to recent surveys and agent benchmarks
\citep{wen2025know,liu2026agentabstain}. We inherit this machinery --- Equation~\eqref{eq:voi}
is its standard stopping-value form --- and do not claim it. What differs here is the
setting in which it is applied: multiple premises per action, verifiers with overlapping
coverage and differing reliability, and adaptive choice of the next check.

\section{Conclusion}
\label{sec:conclusion}

An agent about to act needs to know whether a belief is reliable enough for the specific
thing it is about to do, and if not, whether finding out is worth it. Scoring verification
solely by the expected reduction in the loss of an unchanged action yields no information
benefit under Bayesian updating, because information does not change the average loss of an
action already chosen; verification is worth paying
for exactly when it can change what the agent does next. The current evidence supports two
findings: an action's consequences change how much
verification its premises warrant, and stopping-based VOI removes unnecessary checks --- on ToolSandbox,
\textsc{AMV-Guard} lowers verification cost by 33.9\% relative to a static policy without
losing safe-proceed recall. Richer multi-premise dependency structure, longer verification
sequences, and full cross-environment transfer of the structured pipeline remain to be
validated. We shift the question from \emph{is this memory valid} to \emph{is this memory
valid enough to act on now}.

\section*{Limitations}
\label{sec:limitations}

Premise extraction remains partly model-driven: the counterfactual procedure of
Section~\ref{sec:interventions} filters proposed premises but cannot recover one never
proposed, so extractor recall upper-bounds the pipeline; RQ4 measures this under several
OOD strata. The deployed planner is a bounded approximation: it prefers any positive
one-step VOI and invokes only a depth-two fallback when no such query exists, so it need not
match the full exact planner on longer complementary sequences. The current ToolSandbox
observation model uses binary support/refute outcomes with stated verifier reliability.
A verifier covering several premises produces one shared binary observation: its likelihood
uses the mean covered memory-error probability, the same observation updates all covered
premises, and the live simulator generates it from the first covered premise. This is an
approximation rather than an exact joint observation model.
Likewise, the present risk engine does not execute the oracle's conjunction-only hyperedge
semantics: it couples all jointly marked premises through one shared latent draw but charges
loss when any represented loss group fails. It therefore cannot represent several distinct
joint hyperedges in one candidate.

The loss function requires an executable consequence signal or an explicit policy
specification; where neither exists the asymmetry must be supplied by hand, and sensitivity
analysis characterises but does not remove that dependence. Our simulators do not capture
every real irreversible consequence --- reputational damage, downstream human cost, and
delayed effects are outside what the evaluators score --- so reported harm rates are harm
\emph{on the modelled consequences}. Finally, our oracle guarantees are scoped as in
Appendix~\ref{sec:oracle-limits}. The formulation itself is independent of any particular
memory architecture, planner, or verifier implementation. We also do not yet report
per-perturbation-family and per-action-class breakdowns or scalability over premise count,
verifier count, and dependency depth.

\section*{Ethical Considerations}

\textsc{AMV-Guard} is a safety mechanism: it decides when an agent should re-check a
remembered fact, or decline, before an action with external consequences. The main risk is
over-reliance. Its harm rates are measured only on the consequences the simulators score
(see Limitations), and a deployment that reads a low Gate-harmful rate as a
general safety guarantee could skip human oversight of irreversible actions. The loss
function encodes whose costs count and how much. When it is supplied by hand, it should be
documented and audited, because a mis-specified asymmetry silently moves the boundary
between proceeding and declining. A gate that declines too often also has a cost, which
falls on users whose legitimate requests go unserved. For that reason we report coverage
next to every safety metric.

All experiments use simulated tool environments (ToolSandbox and AppWorld) with synthetic
users, contacts, and accounts. No human participants were recruited, and no personal data
was collected or annotated. We use both benchmarks for research evaluation only. Artifact
licenses, access conditions, and the intended scope of the released derivatives are detailed
in Appendices~\ref{app:licenses} and~\ref{app:intended-use}.

\bibliography{refs}

\appendix

\section{Formulation Details and Proof of Proposition~\ref{prop:regimes}}
\label{app:proof}

\subsection{Execution screens and the cost-sensitive objective}
\label{app:formulation}

\begin{remark}[Which formulation the threshold belongs to]
\label{rem:formulation}
Under the \emph{constrained} reading --- minimise verification cost subject to the
candidate satisfying the execution constraints below --- one stops as soon as execution is
feasible. Under the cost-sensitive objective used here, this is not generally true: a cheap
verification may still lower expected terminal loss by more than $\lambda C(q)$. We
therefore use the execution thresholds only to determine whether \PROCEED{} is admissible;
VOI compares the resulting stopping decision with the cost of acquiring more information.
\end{remark}

\subsection{Consequences of Proposition~\ref{prop:regimes}}

\noindent Three consequences, each a design constraint rather than a curiosity. Below
$\kappa/p^{-}$ verification has \emph{exactly} zero gross value --- the mechanism of
Remark~\ref{rem:degenerate}, since $\Ecal_o[\Vstop(b^{q,o})]=\theta p$ while executing
stays optimal under every observation --- and below $\theta_1$ it is not worth its cost, so
verifying on memory-side uncertainty alone buys nothing. The regimes are traversed by varying $\theta$ with the memory held fixed,
which is the quantitative form of Figure~\ref{fig:motivation}: a score computed from $m$
and $G_M$ alone is constant in $\theta$ while the optimal decision is not, so no such score
can reproduce it. With an imperfect verifier there is a consequence level past which even
a confirming observation cannot license execution, so the rational act is to decline
\emph{without} verifying. With a perfect verifier, verification remains optimal at
arbitrarily high consequence when $c<(1-p)\kappa$, the avoidable abstention loss; otherwise,
verification is never strictly preferred.
Verification demand is therefore not generally monotone in consequence --- it can rise,
then collapse --- which is why \ABSTAIN\ must be a
terminal decision the policy can reach directly rather than a fallback after the budget is
spent, and why verifier reliability must be modelled rather than assumed away.

\subsection{Proof}

Write $\pi=\Pr(\mathrm{confirm})=(1-p)r+p(1-r)$, so $1-\pi=(1-p)(1-r)+pr$, and let
$p^{+},p^{-}$ be as in the statement. Terminal losses are $\theta$ for executing under a
false premise, $0$ for executing under a true one, and $\kappa$ for abstaining, so
$\Vstop(b)=\min\{\theta p,\kappa\}$ and the value of verifying once and then stopping is
\begin{multline*}
V_{\mathrm{ver}}(\theta)=c+\pi\min\{\theta p^{+},\kappa\}\\
+(1-\pi)\min\{\theta p^{-},\kappa\}.
\end{multline*}

\paragraph{Step 1: $p^{+}<p<p^{-}$.}
$p^{+}/p=(1-r)/\pi$ with $\pi=r+p-2pr$, so $p^{+}<p$ iff $1-r<r+p-2pr$, i.e.\ iff
$(1-2r)(1-p)<0$, which holds since $r>\tfrac12$ and $p<1$. Symmetrically
$p^{-}/p=r/(1-\pi)$, and $p^{-}>p$ iff $r(1-p)>(1-p)(1-r)$, i.e.\ iff $r>\tfrac12$.

\paragraph{Step 2: verification is worthless outside $[\kappa/p^{-},\,\kappa/p^{+}]$.}
If $\theta\le\kappa/p^{-}$ then $\theta p^{+}<\theta p<\theta p^{-}\le\kappa$, every
minimum is attained by its $\theta$ term, and the tower property gives
$\pi\theta p^{+}+(1-\pi)\theta p^{-}=\theta p$. Hence
$V_{\mathrm{ver}}-\Vstop=c\ge 0$: verifying costs $c$ and returns nothing, and since
$\theta p\le\kappa$ the optimal terminal decision is to execute. If instead
$\theta\ge\kappa/p^{+}$ then $\theta p^{+}\ge\kappa$ and a fortiori $\theta p^{-}\ge\kappa$
and $\theta p\ge\kappa$, so every minimum equals $\kappa$ and again
$V_{\mathrm{ver}}-\Vstop=c\ge 0$, with abstention now the optimal terminal decision. This is
the substantive half: past $\kappa/p^{+}$ no observation this verifier can return leaves the
agent willing to execute, so paying for one is strictly wasteful.

\paragraph{Step 3: the middle region and its cost condition.}
For $\kappa/p\le\theta<\kappa/p^{+}$ we have $\Vstop=\kappa$ and
$\min\{\theta p^{-},\kappa\}=\kappa$, so
$\Vstop-V_{\mathrm{ver}}=\pi(\kappa-\theta p^{+})-c$, and verification is optimal exactly
when $c<\pi(\kappa-\theta p^{+})$. The right-hand side is positive and decreasing in
$\theta$ on this range, vanishing at $\theta=\kappa/p^{+}$, so the region is an interval
that shrinks as $c$ grows and is empty for $c$ large enough. Together with Step 2 this
gives thresholds $\theta_1\le\theta_2$ with the stated bracketing.

\paragraph{Step 4: reliability and cost determine the upper regime.}
For $r<1$ we have $p^{+}>0$ and therefore
$\theta_2\le\kappa/p^{+}<\infty$. If $r=1$, then $p^{+}=0$,
$\pi=1-p$, and Step 2's reliability-induced upper cutoff disappears. In the region where
abstention is the current stop, verification has value precisely when
$c<\pi\kappa=(1-p)\kappa$, independent of $\theta$. Hence $\theta_2=\infty$ only under
that strict cost condition. If $c\ge(1-p)\kappa$, verification cannot beat direct
abstention at high consequence and the verification interval is empty. \hfill$\square$

\section{Method Details}
\label{app:method}

\paragraph{Premise graph and joint premises.}
The extraction stage retains a graph linking claims, premises, the candidate, and verifier
coverage; the counterfactual oracle additionally records conjunction-only hyperedges when
two flips change the outcome although neither flip does alone. The current deployable risk
engine does not consume distinct hyperedge identifiers. It preserves verifier-to-premise
coverage sets, but represents a multi-premise verifier by one shared binary observation:
its outcome likelihood is approximated from the covered premise beliefs, and the same
observation updates every covered premise. Premises tagged joint share one latent draw, and
loss is charged when any represented loss group fails. We therefore evaluate extractor
layers and consequence conditioning in Section~\ref{sec:rq2}, and treat exact
conjunction-only loss over multiple hyperedges as future work rather than claiming a
hypergraph-versus-flat result.

\paragraph{From memory error to precondition violation.}
The calibrated quantity is memory-claim
error probability, which we convert into action-precondition violation probability before
risk evaluation. If the stored value satisfies the required predicate, violation probability
equals memory-error probability; if it contradicts the predicate, violation probability is
its complement. Premises without a public required value retain the original error
semantics. The two probabilities remain separate so verifier observations update memory
correctness rather than predicate satisfaction directly. These violation probabilities
induce a belief over $z$, and $L$ turns it into the stopping value of
Equation~\eqref{eq:vstop}. Premises assigned to the same loss group represent one downstream
failure event: the group contributes the maximum active impact once, rather than summing
the impact of every member. The asymmetry lives entirely in $L$. The
current ToolSandbox instantiation maps premises to TaskFailure, WrongRecipient,
PolicyViolation, or OtherFailure and fits the first three weights from the training split.
Its reported low/medium/high intervention scales total action consequence; it is not a
separate sensitivity analysis for irreversibility or collateral-damage categories.

\paragraph{Invalidation and replanning.}
There is a separate runtime transition after an observation. If the posterior probability
that a required premise is violated reaches the invalidation threshold $\rho$, the Gate
does not execute the stale candidate and does not treat a replanned action as a terminal
choice inside the same VOI problem. It returns \textsc{ReplanRequired} with the changed
premise and evidence. The agent then either determines that the task is already complete or
submits a new candidate, which is evaluated as a fresh Gate entry using the updated
temporary session memory. The immutable experimental hidden state is not rewritten.

\paragraph{Computational cost.}
\label{sec:complexity}
Repairing the objective is not free. With $m$ affordable verifiers and at most $k$
observations per verifier, one-step selection requires $O(mk)$ posterior stopping-value
evaluations. The two-step fallback requires at most $O(m^2k^2)$ such evaluations, but is
activated only when every one-step score is non-positive. Exact dynamic programming grows
with the full verification decision tree and is therefore reserved for evaluation. Tool
cost remains the primary experimental cost metric; computational search cost is reported
separately when measured rather than inferred from the number of calls.

\subsection{Full Gate algorithm}
\label{app:alg}
\begin{algorithm*}[t]
\caption{\textsc{AMV-Guard} at a candidate commit}
\label{alg:main}
\begin{algorithmic}[1]
\Require candidate $a$, session memory $G_S$, verifiers $Q$, remaining episode budget $B$,
         $\delta,\epsilon,\kappa,\rho,\lambda$
\If{$\mathrm{class}(a) \neq \textsc{Commit}$} \Return $\mathrm{execute}(a)$ \EndIf
\Loop
  \State $P \gets \textsc{ExtractPremises}(a,\mathrm{plan},G_S,\mathrm{schema})$
  \State $G_P \gets \textsc{BuildPremiseGraph}(P,G_S,Q)$
  \State $b_m \gets \textsc{CalibrateMemoryError}(G_P)$;
         $b \gets \textsc{PreconditionViolationBelief}(b_m,P)$
  \State $U\gets\emptyset$ \Comment{verifiers used for the current candidate}
  \State $E\gets\emptyset$
  \Loop
  \If{$\max_{p\in P_{\mathrm{required}}}\Pr_b(p\text{ violated}\mid E)\ge\rho$}
      \State $r\gets\textsc{ContinueAgent}(\textsc{StructuredFeedback}(P,b,E))$
      \If{$r$ completes the task without a commit} \Return \textsc{ResolvedWithoutAction} \EndIf
      \If{$r$ submits a new candidate $a'$}
          \State $a\gets a'$; \textbf{continue outer loop with $B$ unchanged}
      \EndIf
      \State \Return \ABSTAIN
  \EndIf
  \If{$P$ contains \textsc{UnknownUnverifiable}} \State \Return \ABSTAIN \EndIf
  \If{$P$ has an unresolved required conflicting-verifier target}
      \State \Return \ABSTAIN
  \EndIf
  \State $A \gets \{q\in Q\setminus U:q\text{ available},\;r_q>0.5,\;
         \textsc{Covers}(q,P),\;C(q)\le B\}$
  \State $d_{\mathrm{stop}}\gets\textsc{TerminalDecision}(b,a,\delta,\epsilon,\kappa)$
  \State compute $\mathrm{VOI}_1(q)$ for every $q\in A$
  \If{$A\ne\emptyset$ and $\max_{q\in A}\mathrm{VOI}_1(q)>0$}
      \State $q^*\gets\arg\max_{q\in A}\mathrm{VOI}_1(q)$
  \Else
      \State compute $\mathrm{VOI}_2(q)$ for every $q\in A$ if two-step fallback is enabled
      \If{$A\ne\emptyset$ and $\max_{q\in A}\mathrm{VOI}_2(q)>0$}
          \State $q^*\gets\arg\max_{q\in A}\mathrm{VOI}_2(q)$
      \ElsIf{$d_{\mathrm{stop}}=\PROCEED$}
          \State $y\gets\mathrm{execute}(a)$; update session state and $G_S$ from $y$
          \State $r\gets\textsc{ContinueAgent}(y)$
          \If{$r$ submits a new candidate $a'$}
              \State $a\gets a'$; \textbf{continue outer loop with $B$ unchanged}
          \EndIf
          \State \Return the Agent's terminal task outcome
      \Else
          \State \Return \ABSTAIN
      \EndIf
  \EndIf
  \State $o\gets\mathrm{run}(q^*)$ as one shared binary observation over its coverage set
  \State $b_m\gets\mathrm{Bayes}(b_m,q^*,o)$;
         $b\gets\textsc{PreconditionViolationBelief}(b_m,P)$;
         $E\gets E\cup\{(q^*,o)\}$
  \State $\textsc{UpdateSessionMemory}(G_S,o)$;\quad
         $B\gets B-C(q^*)$;\quad $U\gets U\cup\{q^*\}$
  \EndLoop
\EndLoop
\end{algorithmic}
\end{algorithm*}

\subsection{Memory-error probability features}
\label{app:features}
The base calibrator uses numeric features for claim age, a legacy
\texttt{last\_verified\_age\_steps} field that is equal to claim age in this dataset,
source trust, support and conflict counts, a binary provenance-edge indicator stored under
the legacy name \texttt{provenance\_depth}, downstream-premise count, the complement of the
extractor's merged raw confidence, and whether the value was refreshed in the current
runtime. Rule-derived confidence originates in the supporting claim; LLM-derived confidence
is self-reported, and the merged premise keeps the larger value. Categorical
features encode attribute, source type, dependency type, action category, and source layer.
High-cardinality entity identifiers, action targets, scenario families, and perturbation
identities are excluded so the model cannot memorise an instance family. A logistic base
model is fit on training data; leave-one-family-out predictions are used to compare Platt,
beta, and bounded-isotonic calibration by worst-family NLL, overall NLL, and ECE. The
selected probability is clipped to $[0.01,0.99]$ before Gate evaluation.

\section{Protocol Details}

\subsection{Perturbation families}
\label{app:perturbations}
Table~\ref{tab:perturbations} gives the implemented V0--V9 matrix. Messaging supports all
ten families. The three settings families support V0, V1, V2, and V5--V9; V3/V4 are omitted
because their low-battery/service coupling does not provide two independently mutable
critical fields without violating environment invariants. ``Intended response'' is a
design target, not an oracle label; Section~\ref{sec:interventions} computes labels from executable outcomes.

\begin{table*}[t]
\caption{Implemented ToolSandbox perturbations. Counts are train/development/test.}
\label{tab:perturbations}
\centering
\scriptsize
\setlength{\tabcolsep}{3pt}
\begin{tabular}{lp{0.29\textwidth}p{0.20\textwidth}cp{0.28\textwidth}}
\toprule
ID & Construction & Intended response & Count & Automated consistency check \\
\midrule
V0 & No world change. & Proceed when the remembered critical fields agree with the world. & 240/80/80 & Critical memory--world agreement. \\
V1 & Flip exactly one critical field; low-battery is excluded when its cascade would change several services. & Verify if worthwhile; reject/replan if the premise is refuted. & 240/80/80 & Exactly one intended change enters the oracle dependency details. \\
V2 & Flip one causally irrelevant field; fields whose invariants cascade into a critical service are excluded. & Proceed without unnecessary verification. & 240/80/80 & No critical field changes and the selected field is absent from the dependency closure. \\
V3 & Flip two independently sampled critical fields (messaging only). & Exercise multi-premise/adaptive verification. & 160/0/0 & Messaging-only scope and two materialised critical changes. \\
V4 & Apply both fields of one correlated event (contact sync; messaging only). & Exercise correlated-premise handling. & 160/0/0 & Messaging-only scope and both correlated fields materialised. \\
V5 & Flip one critical field and mark every covering channel unavailable/failed ($r=0$). & Abstain when the unsafe premise cannot be checked. & 240/80/80 & Unavailable channels use the failed model and explicit reason. \\
V6 & Flip one critical field and set its covering verifier reliability to $0.9$ or $0.7$. & Verify only when noisy evidence has positive decision value. & 240/80/80 & Reliability value and model label agree; noise seed is recorded. \\
V7 & Flip one critical field and add an equally claimed channel returning the opposing observation. & Abstain unless another channel resolves the conflict. & 240/80/80 & Both channels cover the same target and are marked conflicting. \\
V8 & Flip one critical field; offer a cheap noisy channel ($c=1,r=0.7$) and one reliable physical call with cost units $c=3,r=1$. & Select by reliability-aware net VOI, not call count alone. & 240/80/80 & Expensive channel remains one physical call; cost formula and covers are explicit. \\
V9 & Either lower trust to $0.3$ while leaving a claim correct, or keep trust $0.95$ while making the world stale. & Calibrate from evidence rather than trust alone. & 240/80/80 & Trust case, world-change flag, and changed field are recorded. \\
\bottomrule
\end{tabular}
\end{table*}

Across all rows the manifest additionally checks the configured family/variant/seed
cardinality, unique instance IDs, file hashes, deterministic environment replay from snapshot and seed,
explicit verifier coverage/cost/reliability, resumable candidate transcripts, complete
oracle-labelling outputs (Section~\ref{sec:protocol}), and candidate-level natural-verification bookkeeping. The resulting
3,520 instances split as 2,240/640/640.

\subsection{What the oracles do not certify}
\label{sec:oracle-limits}
A deterministic simulator guarantees that an execution replays. It does not guarantee
that premise enumeration is complete, that an intervened state is internally legal, that
joint dependencies were enumerated rather than sampled, or that the observation model is
correctly specified. Within the finite predicted-premise set, binary observation model,
verifier set, and budget used by the experiment, memoized dynamic programming computes the
exact policy for that model. It does not certify premise completeness, calibration
correctness, or observation-model correctness.

\section{Baseline Roster and Adaptation Rules}
\label{app:baselines}
Always Proceed accepts every candidate without verification. Always Verify repeatedly calls
the cheapest remaining verifier that is available, reliable ($r>0.5$), covers at least one
candidate premise, and fits the remaining episode budget. When no such verifier remains, it
abstains for unknown or uncovered premises and otherwise proceeds; it does not apply
\textsc{AMV-Guard}'s expected-loss/harm terminal screen.
EnvProbe-Simple is the ToolSandbox environment-probing adaptation; AppWorld uses the
corresponding native-agent adaptation. CTA is the cost-aware acquisition baseline under the
shared verifier API and budget. AMV-Static receives the same predicted dependencies,
calibrated beliefs, costs, and reliabilities as the main method but commits to its verifier
selection without posterior-dependent reselection. ExactPlanner-PredictedDep receives those
same non-privileged inputs and solves Equation~\eqref{eq:bellman}; it is an evaluation
reference, not a deployable baseline. Oracle-dependency and oracle-belief substitutions are
reported only as diagnostics. All thresholds are selected on development data and then
frozen for test.

\section{Extended Results}
\label{app:results}

\begin{figure}[h]
\centering
\includegraphics[width=\columnwidth]{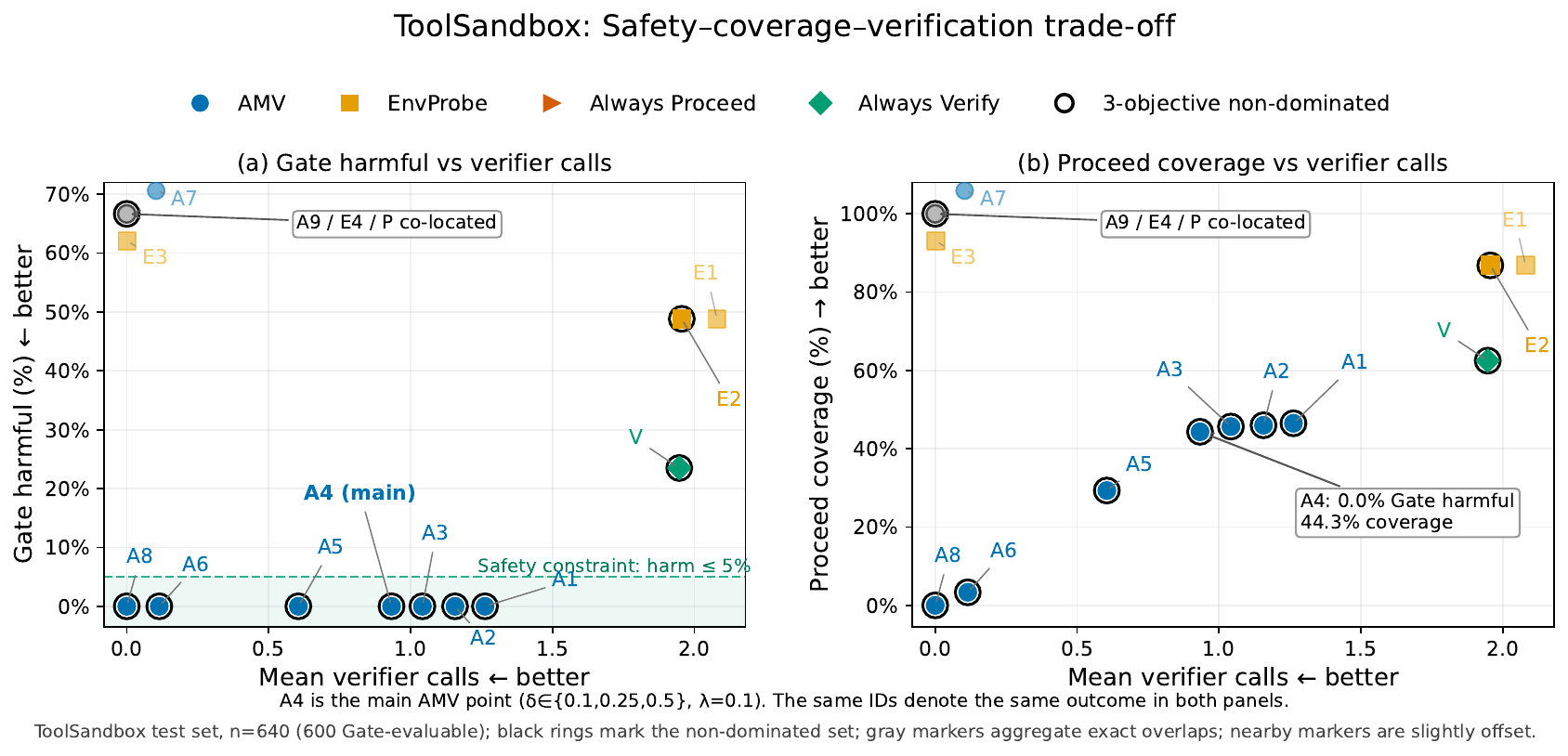}
\caption{ToolSandbox safety--coverage--verification operating points. Left:
Gate harmful against verifier calls. Right: Proceed coverage against verifier calls. Both
rates use the Gate-evaluable denominator, which excludes the 40 test instances with
initial-candidate action or intent errors. The highlighted A4 point has 44.3\% coverage in
the plotted observation seed; Table~\ref{tab:main} aggregates three seeds (44.17\%). A4
merges three settings whose results are identical.}
\label{fig:safety-coverage}
\end{figure}

\begin{figure}[t]
\centering
\includegraphics[width=0.96\linewidth]{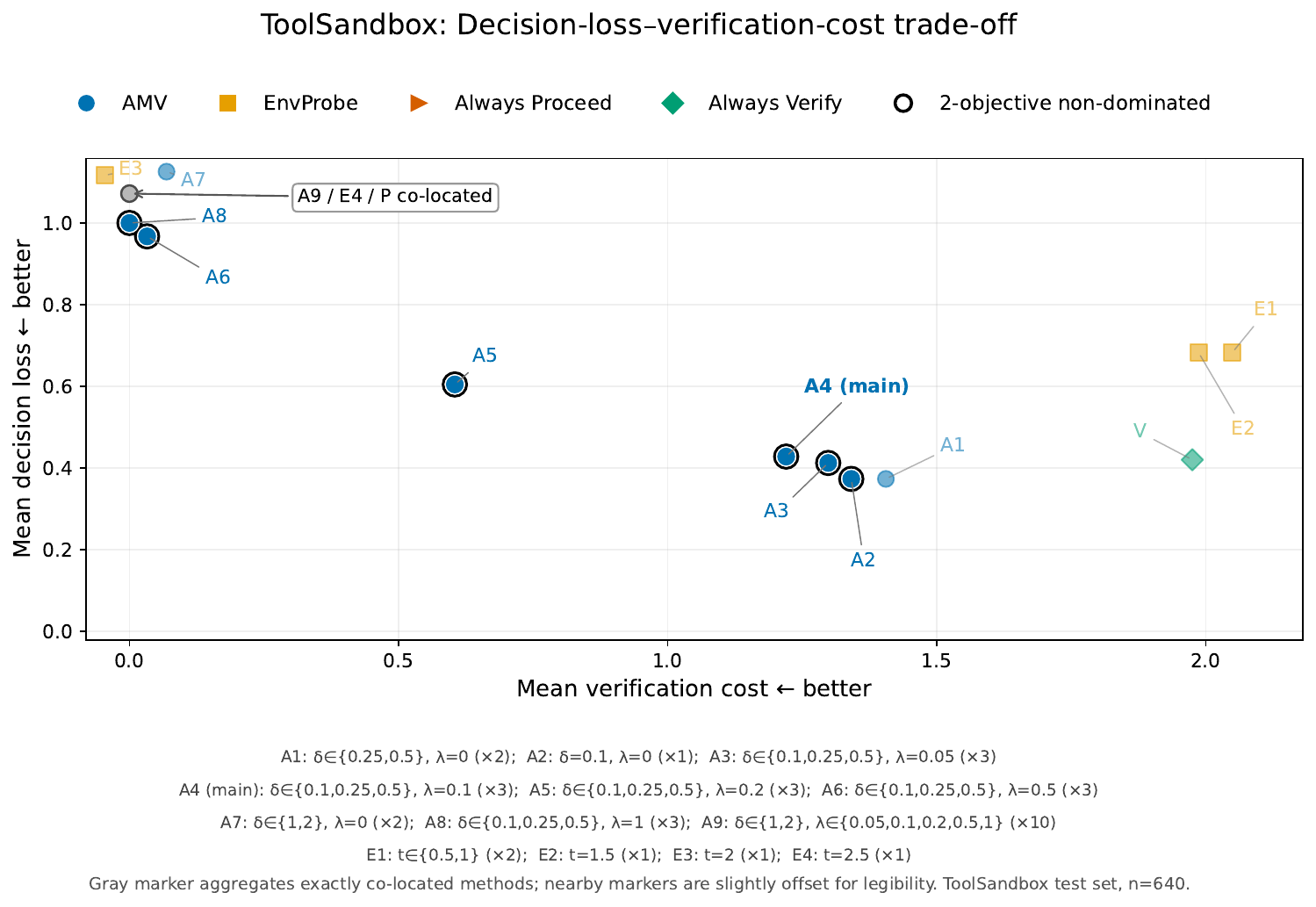}
\caption{ToolSandbox decision-loss--verification-cost operating points over all 37
parameter settings. Settings with identical results are merged in the visualization.}
\label{fig:loss-cost}
\end{figure}

\paragraph{Operating-point sweeps and cross-backbone summary.}
Figure~\ref{fig:safety-coverage} plots Gate harmful and Proceed coverage against verifier
calls, and Figure~\ref{fig:loss-cost} decision loss against verification cost.
Table~\ref{tab:cross-model-compact} summarises the cross-backbone comparison of
Section~\ref{sec:rq1}.

\begin{table}[t]
\caption{Cross-backbone ToolSandbox summary. Each entry is AMV-Static
$\rightarrow$ \textsc{AMV-Guard}; Gate harmful and Coverage use all Gate-evaluable instances.
Each backbone uses its own frozen candidates and train/development calibration.
Bold marks the better of the two (both if tied).}
\label{tab:cross-model-compact}
\centering
\scriptsize
\resizebox{\columnwidth}{!}{%
\begin{tabular}{lccc}
\toprule
Backbone & Gate harmful $\downarrow$ & Verify cost $\downarrow$ & Coverage $\uparrow$ \\
\midrule
Qwen3.5-35B-A3B       & \textbf{0.00\%} $\rightarrow$ \textbf{0.00\%} & 1.843 $\rightarrow$ \textbf{1.219} & 38.83\% $\rightarrow$ \textbf{44.17\%} \\
Qwen3.5-27B           & \textbf{0.00\%} $\rightarrow$ \textbf{0.00\%} & 1.761 $\rightarrow$ \textbf{1.016} & \textbf{40.21\%} $\rightarrow$ \textbf{40.21\%} \\
DeepSeek-V4.1-Flash   & \textbf{0.00\%} $\rightarrow$ \textbf{0.00\%} & 1.916 $\rightarrow$ \textbf{0.787} & \textbf{45.31\%} $\rightarrow$ 39.63\% \\
\bottomrule
\end{tabular}}
\end{table}

\begin{figure}[t]
\centering
\includegraphics[width=0.96\linewidth]{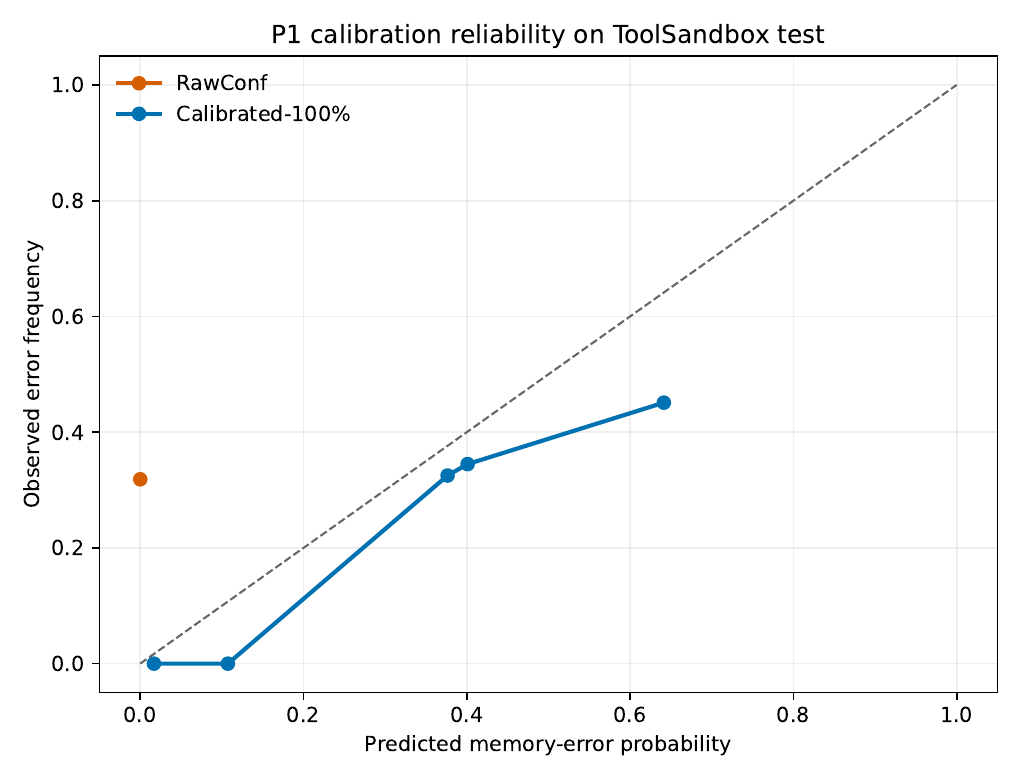}
\caption{Reliability diagrams for raw confidence and for calibrated memory-error beliefs fit
on 100\% of the training data; the other fractions are summarised in
Table~\ref{tab:calibration-data}. Calibration is fit on train only; test is never used for
model or threshold selection.}
\label{fig:calibration-reliability}
\end{figure}

\paragraph{RQ2 tables.} Table~\ref{tab:action-conditioning} gives the action-dependent-premise
intervention, Table~\ref{tab:consequence} the full action-consequence intervention,
Table~\ref{tab:calibration-data} and Figure~\ref{fig:calibration-reliability} calibration
data efficiency, and Table~\ref{tab:premise-ood} the premise-extraction OOD ablation. On the
calibration test split ($n=1,240$ premise records), calibrated ECE/Brier/NLL averaged over
five seeds are 0.114/0.230/0.727 with 10\% fitting data, 0.123/0.230/0.697 with 25\%,
0.108/0.218/0.609 with 50\%, and 0.082/0.209/0.588 with 100\%; at 100\% the Gate obtains
100\% safe-proceed recall, 0\% Gate harmful, 33.33\% Proceed coverage, and 1.167 mean
verification cost. The deployed Union+CF extractor
obtains precision/recall/F1 of 100.0/100.0/100.0 on seen schemas, 100.0/95.91/97.91 on unseen
tool families, 98.66/92.45/95.45 on unseen domain schemas, and 100.0/94.97/97.42 on
unstructured language, with dependency accuracy 100.0\%, 100.0\%, 99.66\%, and 61.92\%.
We do not report a hypergraph-versus-flat runtime ablation because the present risk engine
does not consume distinct hyperedge IDs.

\begin{table*}[t]
\caption{Action-dependent-premise intervention. Both candidates see the same memory graph
and use unit premise impacts and perfect cost-one verifiers. The candidate enters the public
tool-contract rule layer before Gate evaluation. Rates condition on whether the premises
selected for that candidate are the high- or low-risk group.}
\label{tab:action-conditioning}
\centering
\scriptsize
\resizebox{\textwidth}{!}{%
\begin{tabular}{llrrr}
\toprule
Candidate & Extracted premise targets & High-risk verify & Low-risk proceed & Relevant query \\
\midrule
\texttt{send\_message\_with\_phone\_number} & Contact phone; cellular & 100.0 & 100.0 & 100.0 \\
\texttt{set\_wifi\_status} & Wi-Fi; low-battery mode & 100.0 & 100.0 & 100.0 \\
\bottomrule
\end{tabular}}
\end{table*}

\begin{table}[t]
\caption{Action-consequence intervention (500 instances per consequence level). The
candidate action and memory evidence are held fixed while only total loss is rescaled.
This isolates consequence weighting, not distinct-action conditioning. Rates are
percentages; cost is mean verifier cost.}
\label{tab:consequence}
\centering
\scriptsize
\resizebox{\columnwidth}{!}{%
\begin{tabular}{llrrrr}
\toprule
Method & Loss level & Verify & Harmful proceed & Safe completion & Cost \\
\midrule
\multirow{3}{*}{AMV-Guard}
 & Low    & 0.0 & 0.0 & 100.0 & 0.000 \\
 & Medium & 15.8 & 64.2 & 33.6 & 0.164 \\
 & High   & 62.8 & 0.0 & 33.6 & 1.170 \\
\midrule
\multirow{3}{*}{AMV-NoLoss}
 & Low    & 15.8 & 0.0 & 97.8 & 0.164 \\
 & Medium & 15.8 & 64.2 & 33.6 & 0.164 \\
 & High   & 15.8 & 64.2 & 33.6 & 0.164 \\
\midrule
\multirow{3}{*}{EnvProbe-Simple}
 & Low    & 100.0 & 0.0 & 78.2 & 1.000 \\
 & Medium & 100.0 & 45.0 & 33.2 & 1.000 \\
 & High   & 100.0 & 45.0 & 33.2 & 1.000 \\
\midrule
\multirow{3}{*}{Freshness-Only}
 & Low    & 0.0 & 0.0 & 100.0 & 0.000 \\
 & Medium & 0.0 & 66.4 & 33.6 & 0.000 \\
 & High   & 0.0 & 66.4 & 33.6 & 0.000 \\
\midrule
\multirow{3}{*}{Uncertainty-Only}
 & Low    & 100.0 & 0.0 & 61.4 & 1.570 \\
 & Medium & 100.0 & 28.2 & 33.2 & 1.570 \\
 & High   & 100.0 & 28.2 & 33.2 & 1.570 \\
\bottomrule
\end{tabular}}
\end{table}

\begin{table*}[t]
\caption{Calibration data efficiency on the held-out ToolSandbox test split. Calibrated
rows average five instance-grouped seeds. Gate metrics use the corresponding calibrated
beliefs and evaluate a single Gate pass on each of the 600 Gate-evaluable frozen initial
candidates per seed, without replanning: Coverage is the share Proceeded at that initial
Gate, and Cost is the initial-Gate verification cost per candidate. Table~\ref{tab:main}
instead runs full Agent--Gate episodes that include replanning, counts Proceeds after
replanning in Coverage, and averages Cost over all episodes, so the two tables are not
directly comparable (e.g.\ 33.33\% vs.\ 44.17\% coverage for full calibration). All rates are
percentages.}
\label{tab:calibration-data}
\centering
\scriptsize
\resizebox{\textwidth}{!}{%
\begin{tabular}{lrrrrrrr}
\toprule
Fit data & ECE & Brier & NLL & Safe recall & Gate harmful & Coverage & Cost \\
\midrule
Raw confidence & 0.319 & 0.319 & 4.401 & 100.00 & 65.83 & 99.17 & 0.000 \\
10\%  & 0.114 & 0.230 & 0.727 & 86.00 & 3.40 & 32.07 & 0.980 \\
25\%  & 0.123 & 0.230 & 0.697 & 86.00 & 0.00 & 28.67 & 0.998 \\
50\%  & 0.108 & 0.218 & 0.609 & 100.00 & 0.00 & 33.33 & 1.082 \\
100\% & \textbf{0.082} & \textbf{0.209} & \textbf{0.588} & 100.00 & 0.00 & 33.33 & 1.167 \\
\bottomrule
\end{tabular}}
\end{table*}

\begin{table*}[t]
\caption{Premise-extraction OOD ablation. Extraction metrics use all 160 candidates per
stratum; GH and cost use the 158 Gate-evaluable candidates remaining after excluding two
initial action/intent-ineligible cases per stratum. CR is critical premise recall, DA dependency
accuracy, GH Gate harmful, and cost mean verification cost; F1 combines precision and CR.
Union+CF is the deployed
LLM+Graph extractor with counterfactual filtering. All rates are percentages.}
\label{tab:premise-ood}
\centering
\scriptsize
\resizebox{\textwidth}{!}{%
\begin{tabular}{llrrrrrr}
\toprule
Stratum & Extractor & Precision & CR & F1 & DA & GH & Cost \\
\midrule
\multirow{4}{*}{Seen schema}
 & LLM-only & 100.00 & 94.65 & 97.25 & 100.00 & 10.76 & 1.051 \\
 & Graph-only & 100.00 & 100.00 & 100.00 & 62.89 & 0.00 & 0.962 \\
 & LLM+Graph & 100.00 & 100.00 & 100.00 & 100.00 & 0.00 & 0.962 \\
 & Union+CF & 100.00 & 100.00 & 100.00 & 100.00 & 0.00 & 0.962 \\
\midrule
\multirow{4}{*}{Unseen tool family}
 & LLM-only & 100.00 & 95.91 & 97.91 & 100.00 & 8.23 & 1.025 \\
 & Graph-only & 100.00 & 12.58 & 22.35 & 100.00 & 7.59 & 0.000 \\
 & LLM+Graph & 100.00 & 95.91 & 97.91 & 100.00 & 8.23 & 1.025 \\
 & Union+CF & 100.00 & 95.91 & 97.91 & 100.00 & 8.23 & 1.025 \\
\midrule
\multirow{4}{*}{Unseen domain/schema}
 & LLM-only & 98.64 & 91.51 & 94.94 & 98.28 & 17.09 & 1.070 \\
 & Graph-only & 100.00 & 12.58 & 22.35 & 100.00 & 7.59 & 0.000 \\
 & LLM+Graph & 98.66 & 92.45 & 95.45 & 99.66 & 15.19 & 1.051 \\
 & Union+CF & 98.66 & 92.45 & 95.45 & 99.66 & 15.19 & 1.051 \\
\midrule
\multirow{4}{*}{Unstructured language}
 & LLM-only & 100.00 & 94.97 & 97.42 & 61.92 & 10.13 & 1.025 \\
 & Graph-only & 0.00 & 0.00 & 0.00 & 0.00 & 0.00 & 0.000 \\
 & LLM+Graph & 100.00 & 94.97 & 97.42 & 61.92 & 10.13 & 1.025 \\
 & Union+CF & 100.00 & 94.97 & 97.42 & 61.92 & 10.13 & 1.025 \\
\bottomrule
\end{tabular}}
\end{table*}

\paragraph{AppWorld outcomes beyond Proceed.} Besides the 57.37\% Proceed coverage
reported in Section~\ref{sec:rq4}, another 2.11\% of Gate-evaluable AppWorld instances are
Resolved without action; they remain in the coverage denominator without counting as Proceed
or ABSTAIN.

\paragraph{ToolSandbox uncertainty.}
Table~\ref{tab:toolsandbox-ci} gives the mean and 95\% interval across the three frozen
verifier-observation seeds. Intervals are seed-level $t$ intervals clipped to the support of
rate metrics. ExactPlanner-PredictedDep and AMV-Guard coincide at this operating point.

\begin{table*}[t]
\caption{ToolSandbox live end-to-end means with 95\% intervals. Safe recall conditions on
safe, Gate-evaluable initial candidates; Gate harmful and Coverage use all Gate-evaluable
instances (Section~\ref{sec:setup}). Raw harm is end-to-end. Rates are percentages.}
\label{tab:toolsandbox-ci}
\centering
\scriptsize
\resizebox{\textwidth}{!}{%
\begin{tabular}{lrrrrrr}
\toprule
Method & Safe recall & Gate harmful & Verify cost & Task success & Raw harm & Coverage \\
\midrule
Always Proceed & 100.00 [100.00,100.00] & 66.67 [66.67,66.67] & 0.000 [0.000,0.000] & 68.23 [68.01,68.45] & 68.75 [68.75,68.75] & 100.00 [100.00,100.00] \\
Always Verify & 100.00 [100.00,100.00] & 23.67 [23.25,24.08] & 1.976 [1.964,1.988] & 74.58 [73.99,75.18] & 28.02 [27.21,28.83] & 62.67 [62.25,63.08] \\
EnvProbe-Simple & 100.00 [100.00,100.00] & 49.00 [48.59,49.41] & 2.018 [2.016,2.020] & 74.48 [73.89,75.07] & 52.19 [51.80,52.58] & 87.00 [86.59,87.41] \\
CTA & 99.83 [99.12,100.00] & 63.61 [60.48,66.75] & 0.123 [0.123,0.123] & 67.71 [66.24,69.18] & 65.89 [62.95,68.82] & 96.89 [93.52,100.00] \\
AMV-Static & 100.00 [100.00,100.00] & 0.00 [0.00,0.00] & 1.843 [1.831,1.856] & 72.34 [72.34,72.34] & 6.82 [6.60,7.05] & 38.83 [38.83,38.83] \\
AMV-Guard & 100.00 [100.00,100.00] & 0.00 [0.00,0.00] & 1.219 [1.219,1.219] & 73.91 [73.91,73.91] & 7.03 [7.03,7.03] & 44.17 [44.17,44.17] \\
ExactPlanner-PredictedDep & 100.00 [100.00,100.00] & 0.00 [0.00,0.00] & 1.219 [1.219,1.219] & 73.91 [73.91,73.91] & 7.03 [7.03,7.03] & 44.17 [44.17,44.17] \\
\bottomrule
\end{tabular}}
\end{table*}

\paragraph{Cross-backbone ToolSandbox results.}
Table~\ref{tab:cross-model-full} reports the complete six-metric results for every core
method and all three evaluated backbones. Values are means over three verifier-observation
seeds with 640 test episodes per seed. Candidate sets, calibration, and threshold selection
are backbone-specific; metric definitions and the downstream evaluator are shared.

\begin{table*}[t]
\caption{Complete cross-backbone ToolSandbox results. Safe recall conditions on safe
Gate-evaluable initial candidates; Gate harmful and Coverage use all Gate-evaluable
instances. Raw harm is end-to-end. Rates are percentages; verification cost is an episode mean.}
\label{tab:cross-model-full}
\centering
\scriptsize
\resizebox{\textwidth}{!}{%
\begin{tabular}{llrrrrrr}
\toprule
Backbone & Method & Safe recall & Gate harmful & Verify cost & Task success & Raw harm & Coverage \\
\midrule
Qwen3.5-35B-A3B & Always Proceed & 100.00 & 66.67 & 0.000 & 68.23 & 68.75 & 100.00 \\
 & Always Verify & 100.00 & 23.67 & 1.976 & 74.58 & 28.02 & 62.67 \\
 & AMV-Static & 100.00 & 0.00 & 1.843 & 72.34 & 6.82 & 38.83 \\
 & \textsc{AMV-Guard} & 100.00 & 0.00 & 1.219 & 73.91 & 7.03 & 44.17 \\
\midrule
Qwen3.5-27B & Always Proceed & 100.00 & 64.73 & 0.000 & 68.13 & 68.75 & 100.00 \\
 & Always Verify & 100.00 & 23.63 & 2.060 & 75.52 & 30.63 & 63.84 \\
 & AMV-Static & 100.00 & 0.00 & 1.761 & 72.03 & 5.63 & 40.21 \\
 & \textsc{AMV-Guard} & 100.00 & 0.00 & 1.016 & 72.03 & 5.63 & 40.21 \\
\midrule
DeepSeek-V4.1-Flash & Always Proceed & 100.00 & 62.96 & 0.000 & 77.19 & 68.75 & 100.00 \\
 & Always Verify & 100.00 & 19.88 & 2.219 & 79.84 & 28.80 & 65.37 \\
 & AMV-Static & 100.00 & 0.00 & 1.916 & 74.79 & 14.27 & 45.31 \\
 & \textsc{AMV-Guard} & 100.00 & 0.00 & 0.787 & 70.16 & 15.78 & 39.63 \\
\bottomrule
\end{tabular}}
\end{table*}

\paragraph{ToolSandbox secondary outcomes.}
The task-success metric includes both successful proceeds and tasks completed without a
new commit after replanning. Table~\ref{tab:toolsandbox-secondary} makes those outcomes,
initial-candidate action or intent errors, and replan frequency explicit rather than charging
them to the Gate.

\begin{table*}[t]
\caption{ToolSandbox secondary outcomes. Unsafe Pass conditions on harmful, Gate-evaluable
initial candidates. Resolved and Final ABSTAIN use all Gate-evaluable instances and require
no executed Gate-approved action; Final ABSTAIN additionally requires an explicit terminal
ABSTAIN and excludes Resolved without action. Initial candidate action/intent error and Replan
use all episodes. All entries are percentages.}
\label{tab:toolsandbox-secondary}
\centering
\scriptsize
\begin{tabular}{lrrrrr}
\toprule
Method & Unsafe Pass & Resolved w/o action & Final ABSTAIN & Initial action/intent error & Replan \\
\midrule
Always Proceed & 100.00 & 0.00 & 0.00 & 6.25 & 0.00 \\
Always Verify & 35.50 & 22.67 & 14.67 & 6.25 & 17.81 \\
EnvProbe-Simple & 73.50 & 13.00 & 0.00 & 6.25 & 20.63 \\
CTA & 95.42 & 0.11 & 3.00 & 6.25 & 0.00 \\
AMV-Static & 0.00 & 31.83 & 29.33 & 6.25 & 17.08 \\
AMV-Guard & 0.00 & 31.83 & 24.00 & 6.25 & 25.00 \\
ExactPlanner-PredictedDep & 0.00 & 31.83 & 24.00 & 6.25 & 25.00 \\
\bottomrule
\end{tabular}
\end{table*}

\paragraph{AppWorld uncertainty.}
Table~\ref{tab:appworld-ci} reports paired task-bootstrap 95\% intervals (2,000 resamples).
The broad intervals reflect the 100-task transfer set and should temper point-estimate
comparisons between AMV-Guard and AMV-Static.

\begin{table*}[t]
\caption{AppWorld means with paired-bootstrap 95\% intervals. Gate harmful and Coverage use
all Gate-evaluable instances; Unsafe Pass conditions on harmful candidates. Rates are percentages.}
\label{tab:appworld-ci}
\centering
\scriptsize
\resizebox{\textwidth}{!}{%
\begin{tabular}{lrrrrrrr}
\toprule
Method & Safe recall & Gate harmful & Unsafe Pass & Verify cost & Task success & Raw harm & Coverage \\
\midrule
Always Proceed & 100.00 [100.00,100.00] & 18.95 [13.68,23.50] & 37.89 [27.37,47.00] & 0.000 [0.000,0.000] & 35.50 [27.00,44.00] & 19.50 [14.50,24.50] & 100.00 [100.00,100.00] \\
Always Verify & 54.74 [44.68,64.44] & 6.32 [3.16,10.10] & 12.63 [6.32,20.20] & 5.600 [4.280,7.105] & 11.00 [6.00,16.50] & 8.50 [5.00,13.00] & 37.37 [29.79,45.31] \\
EnvProbe & 91.58 [86.02,96.88] & 17.37 [12.89,22.11] & 34.74 [25.77,44.21] & 2.980 [1.965,4.170] & 23.00 [16.00,31.00] & 18.50 [14.00,23.50] & 90.53 [84.90,95.88] \\
CTA & 67.37 [58.33,76.84] & 13.16 [8.95,17.74] & 26.32 [17.89,35.48] & 3.385 [2.600,4.280] & 18.50 [12.50,25.00] & 16.00 [11.00,21.50] & 61.05 [52.69,69.59] \\
AMV-Static & 72.63 [63.16,81.25] & 8.42 [4.74,12.50] & 16.84 [9.47,25.00] & 2.655 [2.095,3.285] & 30.50 [22.00,39.00] & 10.50 [6.50,15.00] & 59.47 [51.03,67.37] \\
AMV-Guard & 70.53 [61.29,79.79] & 7.89 [4.35,11.83] & 15.79 [8.70,23.66] & 2.090 [1.620,2.585] & 28.00 [20.50,36.50] & 10.00 [6.00,14.50] & 57.37 [48.92,65.96] \\
\bottomrule
\end{tabular}}
\end{table*}

\paragraph{Sweep coverage and protocol.}
The loss--cost sweep contains 30 AMV settings from
$\delta\in\{0.1,0.25,0.5,1,2\}$ and
$\lambda\in\{0,0.05,0.1,0.2,0.5,1\}$, five EnvProbe thresholds
($0.5,1,1.5,2,2.5$), and two fixed baselines: 37 settings and 23,680 full episodes in
total. The main ToolSandbox point uses $\delta=0.5$, $\lambda=0.1$, unnecessary
non-proceed penalty $\kappa=1$, invalidation threshold $\rho=0.95$, harm constraint
$\epsilon=0.05$, and an episode-level budget of 4 accumulated across replan. The
action-consequence study contains $5\times3\times500=7,500$ episodes; the main live study
contains $7\times3\times640=13,440$ episodes. All 37 sweep settings passed automated
accounting checks without new model calls; the seven-method main result table passed all
11 metric-accounting checks, and all ten sampled trajectory audits passed.

Calibration is fit on train only, with leave-one-family-out development selection and
10/25/50/100\% data fractions over five instance-grouped seeds. The premise-OOD benchmark
contains 160 distinct candidates in each of four strata (640 candidate IDs total) and four
extractor variants, yielding 2,560 instance-method rows. Extraction metrics use all 160
candidates per stratum; Gate harmful and verification cost use the 158 Gate-evaluable
candidates remaining after excluding two initial action/intent-ineligible cases per stratum.
Distractors appear in every stratum, and oracle dependencies are used only for scoring. AppWorld
uses 100 paired clean/stale tasks, six methods, 1,200 episodes, and paired task-level
bootstrap intervals. Machine-readable per-setting results, calibration rows, and premise
rows will accompany the final artifact package; aggregate tables, audits, fitted
calibrators, and a hash manifest are versioned with the code. No held-out test labels are
used for fitting or threshold selection. The compact file
\texttt{two\_step\_fallback\_dev\_comparison.json} records the two fixed-policy rows from
\texttt{policy\_hyperparameter\_dev.json} at $\kappa=1$, $\lambda=0.1$, and $\rho=0.95$,
including their denominators and the source file's SHA-256.

\section{Reproducibility}
\label{app:repro}
Environment transitions, perturbations, and verifier observations replay deterministically
from snapshots and seeds. External-model outputs are frozen through caches keyed by model,
prompt version, and input hash. Splits are frozen by base scenario before any test-set measurement, and test data is not used to
select features, calibration method, thresholds, or baseline prompts. Small aggregate
tables, audits, configurations, and fitted calibration artifacts are versioned with the
code. We will release the large row-level premise files, per-step decision traces, and
oracle policy trees with the final artifact; the repository manifest records their byte
sizes and SHA-256 hashes. Anonymous repositories are available for the main experiments
(\url{https://anonymous.4open.science/r/AMV-Guard-AEF6/}), the Qwen cross-backbone branch
(\url{https://anonymous.4open.science/r/AMV-Guard-3668/}), and the DeepSeek cross-backbone
branch (\url{https://anonymous.4open.science/r/AMV-Guard-AB3A/}).

\section{License and Terms}
\label{app:licenses}

ToolSandbox is used under Apple's accompanying permissive source-code license
(copyright 2024 Apple Inc.); we retain its copyright, license, and acknowledgement notices.
AppWorld's public code is licensed under Apache License 2.0. Its protected data are subject
to the additional AppWorld condition that public redistribution of the protected portion or
its derivatives remain encrypted. We therefore do not redistribute AppWorld protected data,
derived states, or model caches in plaintext. Our use is limited to the research evaluation
described in this paper and is consistent with the benchmarks' stated research purposes.

The Qwen3.5 model repositories used here are licensed under Apache License 2.0;
the DeepSeek-V4.1-Flash repository is licensed under the MIT License. Inference was performed
through hosted APIs and was additionally governed by the applicable provider terms. We do
not redistribute model weights. With the final artifact, we will release author-written
source code and configurations under Apache License 2.0, and author-generated aggregate
tables and figures under CC BY 4.0. Upstream components and data retain their original
licenses and terms; our release license does not supersede them.

\section{Intended Use and Responsible Release}
\label{app:intended-use}

The released artifact is intended for research on verification policies for tool-using
agents in simulated environments. It is not a production safety certification, a substitute
for human approval of irreversible actions, or a basis for commercial deployment without
independent validation. In particular, the reported harm measures cover only consequences
encoded by ToolSandbox and AppWorld evaluators. The artifact should not be used to infer
protected or sensitive attributes, rank people, or make consequential decisions about real
individuals.

All environment content, user instructions, and tool outputs are in English. All evaluated
users, contacts, accounts, and tool states are simulated; we collected no
human-subject data, personal data, social-media content, or web-scraped user records. This
substantially reduces privacy and re-identification risk, but it also limits external validity:
the simulators do not capture the full distribution of language, accessibility needs,
cultural conventions, or harms faced by marginalized populations. Derived data remain
research-only where required by their source terms. We release aggregate statistics by
default, freeze external-model outputs with provenance metadata, and exclude API credentials
and plaintext protected AppWorld derivatives from the public package.

\section{Models and Computational Budget}
\label{app:compute}

\begin{table*}[t]
\caption{Hosted language models used in the experiments. ``Active'' denotes the approximate
parameters activated per token for mixture-of-experts models; parameter counts are taken from
the official model cards. Model releases: Qwen3.5 \citep{qwen2026qwen35} and
DeepSeek-V4.1-Flash \citep{deepseek2026v41flash}.}
\label{tab:model-compute}
\centering
\small
\begin{tabular}{llll}
\toprule
Model & Total params & Active params & Experimental role \\
\midrule
Qwen3.5-35B-A3B & 35B & 3B & Candidates, premises (\S\ref{sec:protocol}); main, AppWorld (\S\ref{sec:experiments}) \\
Qwen3.5-27B & 27B dense & 27B & Cross-backbone evaluation \\
DeepSeek-V4.1-Flash & 552B & 8B input / 16B output & Cross-backbone evaluation \\
\bottomrule
\end{tabular}
\end{table*}

All foundation-model inference used hosted APIs; no model weights were loaded locally and no
foundation model was trained or fine-tuned. The Qwen models were called through Qwen's
OpenAI-compatible model service under the model IDs \texttt{qwen3.5-35b-a3b} (runs on
10 and 18--19 October 2026) and \texttt{qwen3.5-27b} (12--13 October 2026).
DeepSeek-V4.1-Flash was called through the DeepSeek API under the model ID
\texttt{deepseek-flash} on 12--13 October 2026, after the 10 September 2026 V4.1-Flash
release; the provider's documentation mapped this ID to V4.1-Flash during our runs. The API
providers did not expose accelerator type,
accelerator count, or GPU-hours, so we report requests, tokens, and observed wall-clock time
instead. Local orchestration, calibration, dynamic programming, replay, and aggregation ran
under WSL with Python 3.9 and did not require GPU training.

After deduplicating copied archives and excluding model runs not used in the reported results,
current-model logs identify a directly auditable lower bound of 15,001 API requests. Token
counters are retained for a 7,972-request subset, totaling 9.94 million tokens; the other
7,029 logged live calls do not retain per-call token usage and are therefore excluded from
that token subtotal. Early candidate-generation runs, AppWorld runs, cross-backbone runs,
retries, and discarded development runs predate unified metering. Reconstructing these stages
from episode counts, cached transcripts, and recorded runtimes gives an estimated total of
approximately 72,000 requests and 117 million tokens, with an estimated range of
40,000--100,000 requests and 70--170 million tokens. Retained current-model stage counters
directly measure at least 0.48 hours of parallel wall-clock execution; including unmetered and
repeated runs, we estimate approximately 30 hours in total (roughly 20--45 hours). The latter
figures are estimates under parallel request execution, not summed per-request latency.
Deterministic sweeps over frozen outputs and cached replays incurred no additional model calls.

\section{Software and Implementation Details}
\label{app:software}

The principal environment used Python 3.9.25 with OpenAI Python 2.48.0, NumPy 1.26.4, and
tqdm 4.70.0, together with pandas, SciPy, scikit-learn, NetworkX, and Matplotlib; exact
versions of these packages are recorded in the environment lock file released with the
artifact. ToolSandbox uses Polars 0.20.31. We pin the
ToolSandbox checkout to commit \texttt{165848b9a78cead7ca7fe7c89c688b58e6501219} and the
AppWorld checkout to commit \texttt{42b5bcf3cd334fee33f0c37c02070a9f5807add5}; our adapters
intercept candidate commits, expose verifier observations, and add deterministic snapshot
replay without changing the upstream benchmark evaluators.

External-model calls use deterministic decoding (temperature 0 and top-$p$ 1) and disable
reasoning output where the endpoint supports that option. Stage-specific output limits are
2,048 tokens for candidate generation, 4,096 for dependency extraction, 1,536 for premise
extraction, and 2,200 for the AppWorld agent loop. The memory calibrator uses a standardized
numeric feature block, one-hot categorical features with unknown-category handling, and
balanced logistic regression (maximum 1,000 iterations; fixed random seed); Platt, beta, and
bounded-isotonic alternatives are selected only on development data. Probabilities are
clipped to $[0.01,0.99]$. The complete environment lock file, prompts, model identifiers,
decoding settings, seeds, and input-hash-keyed caches will accompany the final artifact.

\end{document}